\documentclass[aps,prd,showpacs,10pt,nofootinbib]{revtex4-1}

\usepackage{amsfonts,amssymb,amsmath,mathrsfs,latexsym,bbm,dsfont,amsthm}
\usepackage{color}
\usepackage{graphicx}
\usepackage{pstricks,pstricks-add}
\usepackage{pst-plot}
\usepackage[scriptsize,nooneline,hang]{caption}
\usepackage[hang,nooneline,scriptsize]{subfigure}
\usepackage{rotating}
\usepackage[all]{xy}

\definecolor{lightgrey}{rgb}{0.95,0.95,0.95}\definecolor{grey}{gray}{0.7}

\newcommand{\tr}{\tilde{r}}
\newcommand{\trh}{\tilde{\rho}}
\newcommand{\Dr}{\Delta_r}
\newcommand{\tDr}{\tilde{\Delta}_r}
\newcommand{\Lp}{L^{\prime}}
\newcommand{\half}{\frac{1}{2}}

\newpsobject{showgrid}{psgrid}{subgriddiv=1,griddots=10,gridlabels=6pt}

\begin{document}

\title{Analytic treatment of complete and incomplete geodesics in Taub--NUT space--times}

\author{Valeria Kagramanova$^1$, Jutta Kunz$^1$, Eva Hackmann$^2$, Claus L\"ammerzahl$^{2,1}$}
\affiliation{
$^1$Institut f\"ur Physik, Universit\"at Oldenburg,
D--26111 Oldenburg, Germany
\\
$^2$ZARM, Universit\"at Bremen, Am Fallturm,
D--28359 Bremen, Germany }

\date\today

\begin{abstract}

We present
the complete set of analytical solutions of the geodesic equation 
in Taub--NUT space--times in terms of the Weierstra{\ss} elliptic functions. 
We systematically study the underlying polynomials 
and characterize the motion of test particles by its zeros. 
Since the presence of the ``Misner string'' in the
Taub--NUT metric has led to different interpretations,
we consider these in terms of the geodesics of the space--time.
In particular, we address the geodesic incompleteness
at the horizons discussed by Misner and Taub~\cite{MisnerTaub69},
and the analytic extension of Miller, Kruskal and Godfrey~\cite{MiKruGo71},
and compare with the Reissner--Nordstr\"om space--time.  

\end{abstract}


\maketitle

\section{Introduction}

The Taub--NUT solution of the vacuum Einstein equations exhibits
intriguing properties. 
The double name of the solution originates from the two parts the
solution consists of, and at the same time highlights
its strange features.
In 1951 Taub~\cite{Taub51} derived a solution valid in the 
``inner'' region ($\Dr<0$, see Eq.~\eqref{metrikNUTdeSitter}), 
which was interpreted as a cosmological model 
(see e.g.~Brill~\cite{Brill64PR} and Wheeler~\cite{Wheeler62}). 
In 1963 independently Newman, Unti and Tamburino (NUT)~\cite{NUT63} 
obtained a metric valid in the ``outer'' region ($\Dr>0$) 
which reduces to the Schwarzschild metric 
in the limit of vanishing NUT parameter. 
Misner~\cite{Misner63} then suggested to consider the Taub space--time
joined analytically to the NUT space--time as a single Taub--NUT space--time.

The gravitomagnetic properties of the Taub--NUT space--time are  
determined
by the nondiagonal part of the metric, which features the
NUT parameter $n$ as a gravitomagnetic charge.
This nondiagonal term implicates a singularity on the 
half-axis $\vartheta=\pi$, termed ``Misner string'',
which is distinct from the ordinary coordinate singularity 
associated with the use of spherical coordinates.
Misner~\cite{Misner63} suggested to evade this singularity,
by introducing a periodic time coordinate and two coordinate patches.
The first coordinate patch covers the northern hemisphere
and the singularity is located along the axis $\vartheta=\pi$,
while the second patch covers the southern hemisphere
with the singularity extending along the axis $\vartheta=0$. 
(This construction is analogous to the way
in which the ``Dirac string'' singularity in the vector potential
of the Abelian magnetic monopole is eliminated.)
A periodic time coordinate is thus the price to be paid 
for a Taub--NUT space--time free of these axial singularities.

Indeed, this {\it{periodic identification}} of the time coordinate
makes the solution look rather problematic
for physical applications because of the 
resulting causality violations.
Noting that every observer at rest in the coordinate system
would move on a closed time-like world-line, Bonnor~\cite{Bonnor69}
was led to suggest an alternative interpretation of the metric,
in order to evade this unsatisfactory property of the Misner interpretation
of the Taub--NUT space--time.
In his approach Bonnor retained the singularity at $\vartheta=\pi$, 
and endowed it with a physical meaning:
he interpreted it as a semi--infinite massless rotating rod. 
In his interpretation
the NUT space--time is then considered
as being created by a spherically symmetric body 
of mass $M$ at the origin 
and by a source of pure angular momentum (associated with the NUT parameter) 
which is uniformly distributed along the  $\vartheta=\pi$--axis. 
Thus the ``Misner string'' is representing a physical singularity. 
Moreover, close to the singular semi-axis there looms a region of
space--time containing closed time-like curves.
Consequently, to obtain a physically more reasonable interpretation of the
Taub-NUT solution, Bonnor was forced to restrict the coordinate
range, as to exclude the singular regions along with the region
with closed time-like curves.
Containing holes,
the resulting manifold is then incomplete in the sence that
not all geodesics can be extended to arbitrarily large values
of their affine parameters \cite{Bonnor69}.

Recently,
Manko and Ruiz~\cite{Manko2005} reconsidered the interpretation of Bonnor 
for the general family of Taub-NUT solutions, 
containing a constant $C$ in the nondiagonal part of the metric,
$g_{t\varphi}\propto 2n(\cos\vartheta+C)$.
This constant then defines the position of the singularity on the 
axis
of the respective Taub-NUT space--time.
For the special cases, when $C=1$ or $C=-1$ 
(as considered in~\cite{NUT63,Bonnor69}) 
there is only a single semi-infinite singularity 
on the upper or lower part of the symmetry axis, respectively. 
All other $C$--values correspond to Taub-NUT solutions
with two semi-infinite singularities.
The general source associated with
these singularities is then interpreted 
as two semi-infinite counter-rotating rods 
(or only a single semi-infinite rotating rod 
in the special cases $|C|=1$) 
with a finite rod (also rotating in general) inbetween them.
The total mass of the NUT solutions is $M$ 
and does not depend on the choice of $C$. In particular, none of the rods yields a divergent contribution to the total mass.
For the choice $C=0$,
the angular momentum of the middle rod $J_2=C n \sqrt{M^2+n^2}$ vanishes,
and the diverging angular momenta $J_1$ and $J_2$ of the semi-infinite rods
cancel each other.
The analysis of Manko and Ruiz is based on the exact calculation of the Komar integrals for the mass and the angular momentum 
of the three rods constituting the NUT source and their total values.
The unattractive properties of the Bonnor interpretation
are thus retained for this general family of Taub-NUT solutions:
the axial singularity and the closed time-like
curves in its vicinity.

Presuming the {\it{periodic identification}} of the time coordinate,
Misner and Taub~\cite{MisnerTaub69} explored
the geodesics of the Taub-NUT space--time 
and detected the presence of incomplete geodesics.
This may seem surprising, since the origin and the
axis are regular in their interpretation of the space--time.
However, employing Eddington-Finkelstein coordinates,
Misner and Taub found a singular behaviour 
of some geodesics at the horizons: 
as one of the horizons is approached for the second time,
the modified time coordinate diverges 
and the affine parameter terminates, 
thus making these geodesics incomplete.
Now the same type of behaviour is found for 
geodesics in the Reissner-Nordstr\"om space--time,
and is dealt with by analytical continuation
of the Reissner-Nordstr\"om space--time at the horizons,
which consequently allows the continuation of these respective geodesics.
The obvious question following from this observation is therefore
whether the Taub-NUT space--time can be analytically
continued in an analogous fashion as the Reissner-Nordstr\"om space--time.

Indeed, in 1971 Miller, Kruskal and Godfrey~\cite{MiKruGo71} applied
Kruskal-like coordinates and provided an analytic extension 
of the Taub-NUT metric for the entire range of $r$--values. 
Consequently, in their approach the geodesics, 
which previously terminated at a horizon,
could in principle be extended in these new coordinates.
They would no longer be incomplete.
However, the analytic extension
introduced in \cite{MiKruGo71}
came at the prize that the periodic identification
of the time coordinate was no longer possible,
since it would violate the Hausdorff property of the manifold.
Consequently, without the periodic identification
the singularities on the axis are not evaded,
and therefore the extension of Miller, Kruskal and Godfrey~\cite{MiKruGo71}
is geodesically incomplete, as they note.

Leaving aside the incompleteness of certain classes of geodesics,
which apparently cannot be avoided in either interpretation
of the vacuum Taub-NUT space--time,
let us now address some further interesting aspects of this
space--time.
First of all we note, that
the geodesics of the Taub-NUT space--time share 
many of the properties of the trajectories of charged particles in the
field of a magnetic monopole. 
A thorough discussion and comparison of these orbits can be found 
in Zimmerman and Shahir~\cite{ZimSha89}. 
One particular such property is, 
that test particles always move on a cone.
This was also pointed out by Lynden-Bell and Nouri-Zonos~\cite{BellZonos98},
who studied a Newtonian analogue of monopole space--times and discussed
the observational possibilities for (gravito)magnetic monopoles. 
In fact, the spectra of supernovae, quasars or AGN might be
good candidates to infer the existence of (gravito)magnetic monopoles.

On the other hand, the lack of observational evidence
for the existence of gravitomagnetic masses might be explained 
by its presumably very large value~\cite{MuellerPerry86},
and by not sufficiently sensitive instruments~\cite{RahvarZonos2003MNRAS}. 
Nouri--Zonos and Lynden--Bell~\cite{ZonosBellLensing97} 
suggested to look for gravitomagnetic masses by means of gravitational lensing.
By considering the propagation of light in Taub--NUT space--time 
they showed that the NUT deflector 
influences the gravitational lensing effect on the null geodesics. 
In contrast to the lensing in Schwarzschild space--times,
the gravitomagnetic field shears the observed form of the source. 
The possibility to detect and the method to measure 
the gravitomagnetic masses with the next generation 
of microlensing experiments was studied in~\cite{RahvarHabibi2004}. 
In~\cite{Shen_GRG_2002} even the
anomalous acceleration of the Pioneer spacecraft 
was associated with a gravitomagnetic charge.

We present in this paper the complete set of analytic solutions of the geodesic equation 
in Taub-NUT space--times in terms of the Weierstra{\ss} elliptic functions (the complete set of analytical solutions of the geodesic equation in Pleba\'nski--Demia\'nski space-times which Taub--NUT is a special case of, can be found in \cite{PleDemMeth}). 
We classify the orbits by means of the analysis of the underlying polynomials 
which include the parameters of the metric 
and the physical quantities characterizing the test particles. 
In particular, we discuss the incomplete geodesics,
arising in Eddington--Finkelstein coordinates,
as one of the horizons is approached for the second time,
and we address the analytic extension of the space--time
necessary to extend the geodesics,
but forcing us to retain the singularity on the axis. We also discuss observable effects in Taub--NUT space--times on the basis of analytically defined observables, thus making other approaches to experimental signals of the NUT parameter, e.g. \cite{we_at_cqg,Binietal03}, more rigorous.

\section{The geodesic equation}

The general Taub--NUT solution of the Einstein field equations is described by the metric~\cite{NUT63,Misner63} 
\begin{equation}
ds^2 = \frac{\Dr}{\rho^2} \left(dt - 2 {n} (\cos\vartheta + C) d\varphi\right)^2 - \frac{\rho^2}{\Dr}dr^2 - \rho^2\left( d\vartheta ^2 + \sin^2\vartheta  d\varphi ^2 \right) \, , \label{metrikNUTdeSitter}
\end{equation}
where $\rho^2=r^2+{n}^2$ and $\Dr= r^2 - 2 M r - {n}^2 $. Here $M$ is the (gravitoelectric) mass of the solution, and ${n}$ is the NUT charge, regarded as gravitomagnetic mass. For $n \neq 0$ there are always two horizons, defined by $\Delta_r = 0$, and given by
\begin{equation}
r_\pm = M \pm \sqrt{M^2 + n^2} \, .
\end{equation}
One horizon is located at $r_+ > 2 M$, the other 
at\footnote{From $r_{-} = M - \sqrt{M^2 + n^2}$ we obtain $r_{-}^2 - 2 r_- M = n^2$. Since $r_{-} < 0$ 
it follows that $-|n| < r_-$.}  $- |n| < r_{-} < 0$. Between the horizons the radial coordinate $r$ becomes timelike, and the time coordinate $t$ spacelike. 

For test particles or light ($g_{\mu\nu} u^\mu u^\nu = \delta$,
$\delta = 1$ for massive test particles and $\delta = 0$ for light) 
moving on a geodesic one immediately infers the conserved energy $E$ and angular momentum $L$
\begin{eqnarray}
{E} & = & \frac{\Dr}{\rho^2}\left( \frac{dt}{d\tau} - 2 {n} (\cos\vartheta + C)  \frac{d\varphi }{d\tau} \right) \\ 
L & = & 2 {n} (\cos\vartheta + C) \frac{\Dr}{\rho^2} \left(\frac{dt}{d\tau} - 2 {n} (\cos\vartheta + C)  \frac{d\varphi}{d\tau}\right) + \rho^2\sin^2\vartheta  \frac{d\varphi}{d\tau}   \label{muL} \ ,
\end{eqnarray}
where $\tau$ is an affine parameter along the geodesic.

For convenience, we introduce dimensionless quantities ($r_{\rm S} = 2 M$)
\begin{equation}
\label{normpar}
\tr=\frac{r}{r_{\rm S}} \ , \,\, \tilde{t}=\frac{t}{r_{\rm S}} \ , \,\, \tilde{\tau}=\frac{\tau}{r_{\rm S}} \ , \,\,  \tilde{n}=\frac{{n}}{r_{\rm S}} \ , \,\, \tilde{L}=\frac{L}{r_{\rm S}} \ .
\end{equation}It is straightforward to see that in Taub--NUT space--times the Hamilton--Jacobi equation 
\begin{equation}
2\frac{\partial S}{\partial\tau} = g^{ \mu\nu}\frac{\partial S}{\partial x^\mu}\frac{\partial S}{\partial x^\nu}
\end{equation}
is separable and yields for each coordinate a corresponding differential equation
\begin{eqnarray}
\left(\frac{d\tr}{d\gamma}\right)^2 & = & R \label{eq-r-theta:1} \\
\left(\frac{d\vartheta }{d\gamma }\right)^2 & = & \Theta \,  \label{eq-r-theta:2} \\ 
\frac{d\varphi}{d\gamma} & =  & \frac{\Lp - 2 \tilde{n} {E} \cos\vartheta }{\sin^2\vartheta} \ , \,\,\, \text{with} \,\,\, \Lp=\tilde{L}-2\tilde{n}EC \label{dvarphidgamma} \\
\frac{d \tilde{t}}{d\gamma} & = & {E} \frac{\trh^4}{\tDr} + 2 \tilde{n} (\cos\vartheta + C) \frac{\Lp - 2 \tilde{n} {E} \cos\vartheta}{\sin^2\vartheta} \, .  \label{dtildetdgamma}
\end{eqnarray}
Here we used $\trh^2 = \tr^2 + \tilde{n}^2$ and $\tDr=\tr^2 - \tr - \tilde{n}^2$ and introduced the Mino time $\gamma$ through $\trh^2 d\gamma = d\tilde{\tau}$ \cite{Mino03}. We also defined
\begin{eqnarray}
R & = & (\tr^2 + \tilde{n}^2)^2 {E}^2 - \tDr (\delta \tr^2 + \tilde{L}^2 + {k}) \label{R_polynomial} \\
\Theta & = & {k} - \delta \tilde{n}^2 +\tilde{L}^2 - \frac{(\Lp - 2\tilde{n}{E} \cos\vartheta		 )^2}{\sin^2\vartheta}  \label{Theta_polynomial} \, . 
\end{eqnarray}
The separation constant $k$ is known as Carter constant. 
A study of the geodesic equation in Weyl coordinates has been
performed in \cite{valeria}.

\subsection{Gauge transformation}\label{gauge}

Consider a transformation $t=t^\prime + 2 n C \varphi$. It brings the metric~\eqref{metrikNUTdeSitter} into $C$-independent form:
\begin{equation}
ds^2 = \frac{\Dr}{\rho^2} \left(dt^\prime - 2 {n} \cos\vartheta d\varphi\right)^2 - \frac{\rho^2}{\Dr}dr^2 - \rho^2\left( d\vartheta ^2 + \sin^2\vartheta  d\varphi ^2 \right) \ , \label{metrikNUTdeSitter2}
\end{equation}
with corresponding conserved energy $E^\prime=E$ and angular momentum $\Lp=\tilde{L}-2\tilde{n}EC$ for a test particle or light. Thus, the apt notation $\Lp$ introduced in the Hamilton-Jacobi equations~\eqref{eq-r-theta:1}-\eqref{dtildetdgamma} represents a conserved angular momentum in the gauge transformed metric~\eqref{metrikNUTdeSitter2}.  

The Hamilton-Jacobi equation for each coordinate of the metric~\eqref{metrikNUTdeSitter2} yields
\begin{eqnarray}
\left(\frac{d\tr}{d\gamma}\right)^2 & = & R \label{eq-r-theta:1_2} \\
\left(\frac{d\vartheta }{d\gamma }\right)^2 & = & \Theta \,  \label{eq-r-theta:2_2} \\ 
\frac{d\varphi}{d\gamma} & =  & \frac{\Lp - 2 \tilde{n} {E} \cos\vartheta }{\sin^2\vartheta} \,  \label{dvarphidgamma_2} \\
\frac{d \tilde{t}^\prime}{d\gamma} & = & {E} \frac{\trh^4}{\tDr} + 2 \tilde{n} \cos\vartheta \frac{\Lp - 2 \tilde{n} {E} \cos\vartheta}{\sin^2\vartheta} \, ,  \label{dtildetdgamma_2}
\end{eqnarray}
with
\begin{eqnarray}
R & = & (\tr^2 + \tilde{n}^2)^2 {E}^2 - \tDr (\delta \tr^2 + \Lp{^2} + {k}^\prime) \label{R_polynomial_2} \\
\Theta & = & {k}^\prime - \delta \tilde{n}^2 + \Lp{^2} - \frac{(\Lp - 2\tilde{n}{E} \cos\vartheta		 )^2}{\sin^2\vartheta}  \label{Theta_polynomial_2} \, . 
\end{eqnarray}
The Carter constant $k^\prime$ is related to $k$ as $k^\prime-k=\tilde{L}^2-\Lp{^2}$.

From the considerations above one concludes that the metric~\eqref{metrikNUTdeSitter} can be brought into form~\eqref{metrikNUTdeSitter2} which corresponds to the metric~\eqref{metrikNUTdeSitter} for $C=0$. The same feature is observed for the Hamilton-Jacobi equations~\eqref{eq-r-theta:1}-\eqref{dtildetdgamma} and~\eqref{eq-r-theta:1_2}-\eqref{dtildetdgamma_2} under interchange of the constants $\Lp$ and $\tilde L$,  $k^\prime$ and $k$ and the coordinates $\tilde{t}^\prime$ and $\tilde t$.

\section{Complete classification of geodesics}\label{theta-r-features}

Consider the Hamilton-Jacobi equations~\eqref{eq-r-theta:1}-\eqref{dtildetdgamma}. The properties of the orbits are given by the polynomial $R$~\eqref{R_polynomial} and the function $\Theta$~\eqref{Theta_polynomial}. The constants of motion (energy, angular momentum and separation constant) as well as the parameters of the metric (dimensionless NUT--parameter) characterize these polynomials and, as a consequence, the types of orbits. In this section we discuss the motion in Taub--NUT space--times in terms of the properties of the underlying polynomial $R$ and the function $\Theta$. 

\subsection{The $\vartheta$--motion}\label{subsec:theta-pot}

In order to obtain from Eq.~\eqref{eq-r-theta:2} real values 
of the coordinate $\vartheta$ we have to require $\Theta\geq 0$. 
This implies $c_2 := {k}  - \delta  \tilde{n}^2 + \tilde{L}^2 \geq 0 $.
With the new variable $\xi := \cos\vartheta$, Eq.~(\ref{eq-r-theta:2}) 
turns into the equation 
\begin{equation}
\left(\frac{d\xi}{d\gamma}\right)^2 = \Theta_\xi \quad \text{with} \quad \Theta_\xi := a \xi^2 + b \xi + c \, , \label{xieom}
\end{equation}
with a simple polynomial of second order on the right hand side, where $a = - (c_2 + 4\tilde{n}^2E^2)$, $b = 4 \tilde{n} {E} \Lp$, and $c = c_2 - \Lp{^2}$.
Since $ c_2 \geq 0$ we have $a < 0$. This means that $\Theta_\xi$ can be positive if and only if there are real zeros of $\Theta_\xi$. The polynomial $\Theta_\xi$ plays the role of an effective potential for the $\vartheta$--motion. The zeros of $\Theta$ define the angles of two cones which confine the motion of the test particles. 
(A similar feature appears in Kerr space--times~\cite{KerrMeth}.) 
Moreover, every trajectory is not only constrained by these cones but lies itself on a cone in 3--space~\cite{MisnerTaub69,ZimSha89,BellZonos98} (the discussion for $C=-1$ in the papers~\cite{MisnerTaub69,ZimSha89} can be extended for the general family of Taub-NUT solutions).

The discriminant $D = b^2 - 4 a c$ of the polynomial $\Theta_\xi$ 
can be written as $D = 4 c_1 c_2$ with 
$c_1 := c_2 - \Lp{^2} + 4\tilde{n}^2E^2$. 
The existence of real zeros of $\Theta_\xi$ requires $D \geq 0$.
This implies that both $c_1$ and $c_2$ should be non--negative
\begin{equation}
\begin{array}{ll} c_1 = c_2 - \Lp{^2} + 4\tilde{n}^2E^2 & \geq 0 \\
                  c_2 = {k} - \delta  \tilde{n}^2 + \tilde{L}^2  &  \geq 0 \ . \end{array} \  \label{theta-cond-lamu} 
\end{equation}
These are conditions on the parameters $E$, $\tilde{L}$, and $k$ 
for some given $\tilde{n}$. 
As long as ${k} - \delta \tilde{n}^2$ is positive 
there are no constraints on $\tilde{L}$ and ${E}$. 
When ${k} - \delta \tilde{n}^2$ becomes negative 
the inequalities \eqref{theta-cond-lamu} imply lower limits 
for the energy and angular momentum 
given by $\tilde L_{\rm min} = \pm\sqrt{\delta \tilde n^2 - k}$ and $E_{\rm min} = \frac{1}{2 \tilde n (1-C^2)} \left(-\tilde{L}C \pm \sqrt{\delta \tilde n^2 - k + c_2 C^2} \right)$. 

One can show that
\begin{equation}
\Theta = c_1 - \frac{(\Lp \cos\vartheta - 2 E \tilde n)^2}{\sin^2\vartheta} = c_2 - \frac{(\Lp - 2 E \tilde n \cos\vartheta)^2}{\sin^2\vartheta} \, \label{Theta_c1c2} .
\end{equation}
The zeros of $\Theta_\xi$ are given by
\begin{equation}
\xi_{1,2}= \frac{2 E \tilde{n} \Lp \mp \sqrt{c_1c_2}}{c_1 + \Lp{^2}} \ . \label{OpeningAngles}
\end{equation}
The conditions~\eqref{theta-cond-lamu} ensure the compatibility of $\xi \in [-1, 1]$ with $\Theta_\xi \geq 0$.

The function $\Theta_\xi$ describes a parabola with the maximum at $\displaystyle{\left(\frac{2\tilde{n}{E}\Lp}{c_1+\Lp{^2}}, \frac{c_1c_2}{c_1+\Lp{^2}}\right)}$. 
For nonvanishing $\tilde n$, $E$, and $\tilde L$ the maximum of the parabola is no longer located at $\xi = 0$ or, equivalently, the zeros are no longer symmetric with respect to $\xi = 0$. Only for vanishing $\tilde n$, $E$, or $\tilde L$ both cones are symmetric with respect to the equatorial plane. 

The $\vartheta$--motion can be classified according to
the sign of $c_2 - \Lp{^2}$:
\begin{enumerate}  
\item If $c_2 - \Lp{^2} < 0$ then $\Theta_\xi$ has 2 positive zeros for $\Lp {E} \tilde{n} > 0$ and $\vartheta  \in  (0, \pi/2)$, so that the particle moves above the equatorial plane without crossing it. If $\Lp {E} \tilde{n} < 0$ then $\vartheta  \in  (\pi/2, \pi)$.

\item If $c_2 - \Lp{^2} = 0$ then $\Theta_\xi$ has two zeros: $\xi_1=0$ and $\xi_2=\frac{4 {E} \tilde{n} \Lp}{\Lp{^2} + 4 {E}^2 \tilde{n}^2}$. If $\Lp {E} \tilde{n} \geq 0$ then $\xi \in [0, 1)$ and $\vartheta \in (0, \frac{\pi}{2}$]. If $\Lp {E} \tilde{n} \leq 0$ then $\xi \in (-1, 0]$ and $\vartheta \in [\frac{\pi}{2}, \pi)$. If $\Lp=2 {E} \tilde{n}$ 
then the $\vartheta$--motion fills the whole upper hemisphere $\vartheta \in [0, \frac{\pi}{2}]$. The motion fills the whole lower hemisphere with $\vartheta \in [\frac{\pi}{2}, \pi]$ if $\Lp = - 2 {E} \tilde{n}$.

\item If $c_2 - \Lp{^2} > 0$ then $\Theta_\xi$ has a positive and a negative zero and  $\vartheta \in (0,\pi)$,  and the particle crosses the equatorial plane during its motion. 
\end{enumerate}

In general, the second term of the function $\Theta$ 
in \eqref{Theta_c1c2} diverges for $\vartheta \rightarrow 0, \pi$. 
However, if $\Lp = 2 E \tilde n$ this term is regular 
for $\vartheta = 0$ and if $\Lp = - 2 E \tilde n$ it is regular for $\vartheta = \pi$. If $\Lp = \pm 2 E \tilde n$, then $c_1 = c_2$. 
The regularity of $\Theta$ in these cases can be seen from
\begin{equation}
\Theta = c_2 - 4 E^2 \tilde n^2 \frac{(1 \mp \cos\vartheta)^2}{\sin^2\vartheta} \, , 
\end{equation}
by application of L'H\^{o}pital's rule. 
If, furthermore, $E^2 = 0$ then $\Theta = k - \delta \tilde n^2 + \tilde{L}^2$ which is independent of $\vartheta$. 

In the special cases when one of the constants $c_1$ or $c_2$ or both vanish, 
$\Theta_\xi$ has a double root which is the only possible value for 
$\vartheta$. One can distinguish three cases:
\begin{enumerate}
\item If $c_1 = 0$ and $c_2 > 0$, then $\xi = \frac{2{E} \tilde{n}}{\Lp}$ for $ \Lp{^2} > 4E^2n^2 $.
\item If $c_2 = 0$ and $c_1 > 0$ then $\xi = \frac{\Lp}{2{E} \tilde{n}}$ for $ \Lp{^2} < 4E^2n^2 $.
\item If $c_1 = c_2 = 0$ then $\xi = \pm 1$ implying that 
$\vartheta = 0$ or $\vartheta = \pi$ are possible. 
In this case $\Lp = \pm 2 E \tilde n$ (as discussed above).  
\end{enumerate} 
This means that during a test particle's motion the coordinate $\vartheta$ is constant and the trajectory lies on a cone around the $\vartheta=0, \pi$--axis with the opening angle $\arccos\xi$. In this case we immediately obtain from \eqref{dvarphidgamma} that $\varphi(\gamma) = {\cal C} (\gamma - \gamma_{\rm in})$ with a constant 
${\cal C} = \frac{\Lp - 2 \tilde{n} {E} \xi}{1-\xi^2}$ which in case 1 is ${\cal C} = \Lp$ and in case 2 is ${\cal C} = 0$. 

Thus, the nonvanishing of $c_1$ and $c_2$ indicates that the motion of the particle is not symmetric with respect to the $\vartheta=0, \pi$--axis. Therefore, these two constants may be regarded to play the role of a generalized Carter constant which appears, e.g., in the motion of particles in a Kerr space--time.

\subsection{The $\tr$--motion}\label{subsec:r-motion}

\subsubsection{Possible types of orbits}

Before discussing the $\tr$--motion we introduce a list of all possible orbits:
\begin{enumerate}\itemsep=-2pt
	\item {\it Transit orbits} (TO) with $\tr \in (\pm\infty, \mp\infty)$ with a single transit of $r = 0$.
	\item {\it Escape orbits} (EO) with ranges 
\begin{enumerate}
	\item $(r_1, \infty)$ with $r_1 > 0$ or
	\item $(-\infty, r_1^\prime)$ with $r_1^\prime < 0$.
\end{enumerate}
These escape orbits do not cross $r = 0$. 
	\item {\it Crossover escape orbits} (CEO) with ranges 
\begin{enumerate}
	\item $(r_1, \infty)$ with $r_1 < 0$ or
	\item $(-\infty, r_1^\prime)$ with $r_1^\prime > 0$,
\end{enumerate}
which cross $r = 0$ two times. 
	\item {\it Bound orbits} (BO) with range $\tr \in (r_1, r_2)$ with $r_1 < r_2$ and
\begin{enumerate}
	\item either $r_1, r_2 > 0$ or 
	\item $r_1, r_2 < 0$. 
\end{enumerate}
	\item {\it Crossover bound orbits} (CBO) with range $\tr \in (r_1, r_2)$ where $r_1 < 0$ and $r_2 > 0$.  
\end{enumerate}

The mathematical condition from the geodesic equation
for crossing the origin $\tr=0$ is $R(0) \geq 0$. 
This implies 
$\displaystyle{R(0)=\tilde{n}^2 \left(\tilde{n}^2 {E}^2 
+ c_2 + \delta \tilde{n}^2 \right)}\geq0$, 
which is always fulfilled under the conditions~(\ref{theta-cond-lamu}).

\subsubsection{The radial motion}\label{radmotion}

\begin{figure}[th!]
\begin{center}
\subfigure[][$n=0.05, k=0.0$]{\label{nutn005k0}\includegraphics[width=3.9cm]{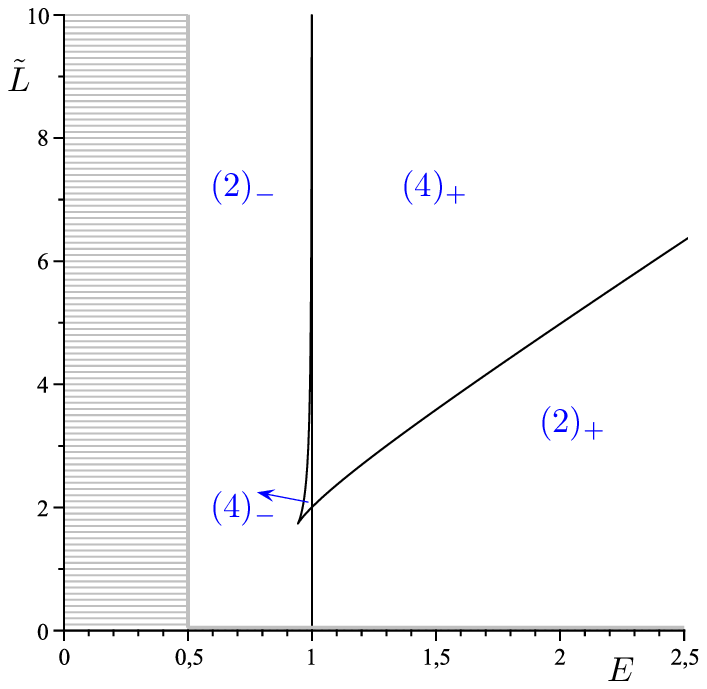}}
\subfigure[][$n=0.5, k=0.0$]{\label{nutn05k0}\includegraphics[width=3.9cm]{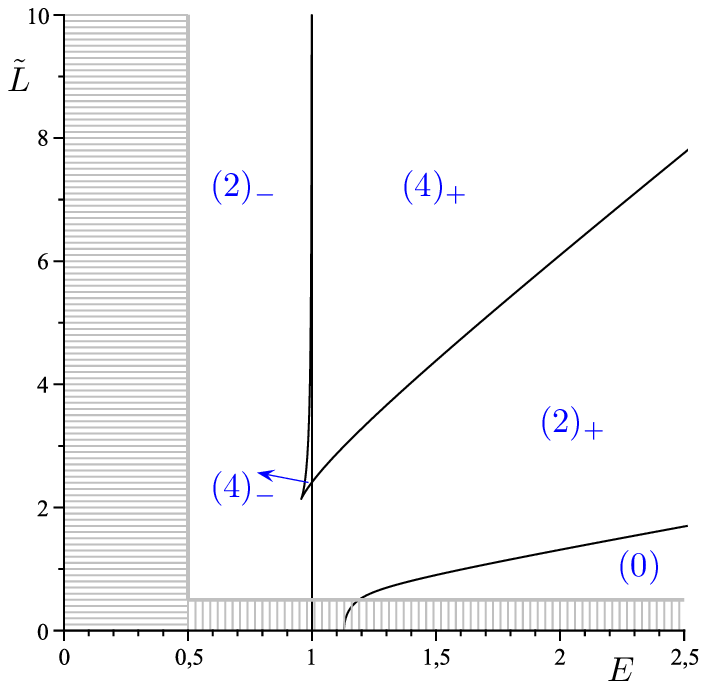}}
\subfigure[][$n=2.0, k=0.0$]{\label{nutn2k0}\includegraphics[width=3.9cm]{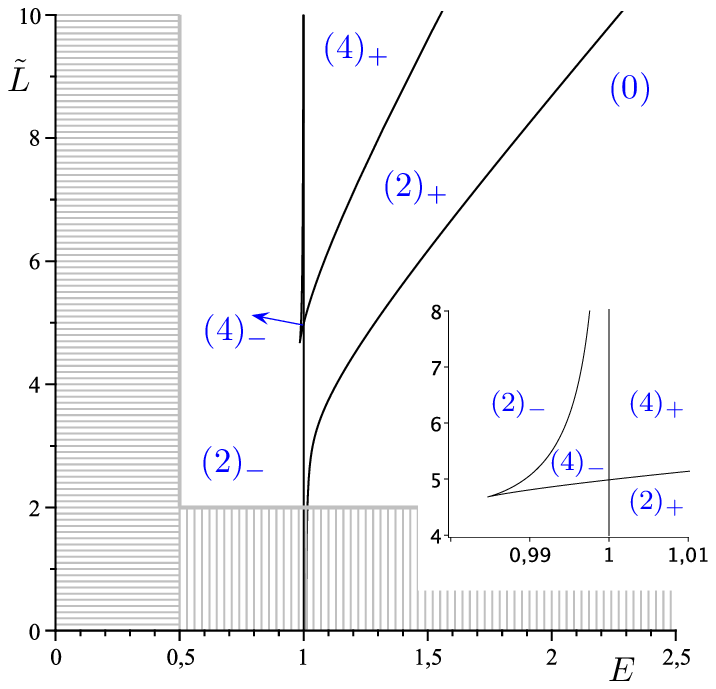}}
\subfigure[][$n=5.0, k=0.0$]{\label{nut5k0}\includegraphics[width=3.9cm]{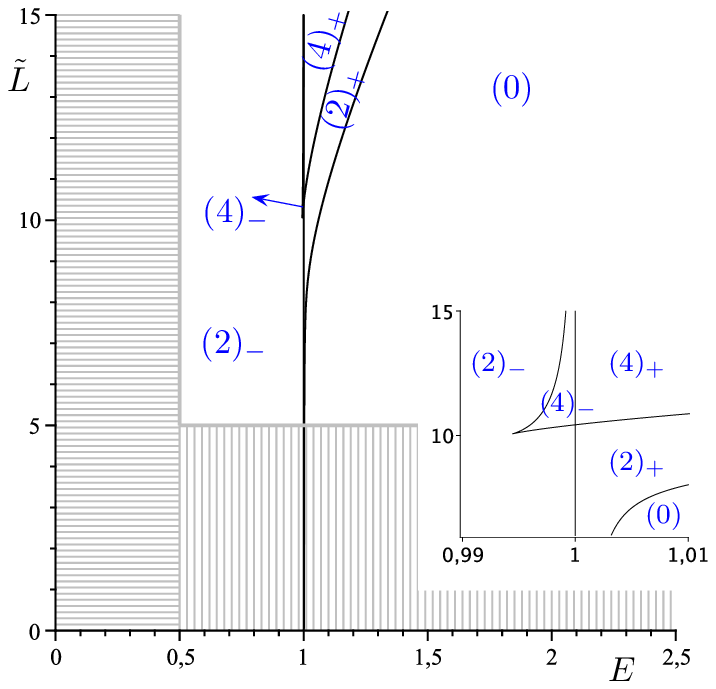}}
\subfigure[][$n=0.5, k=-2.0$]{\label{nutn05k-2}\includegraphics[width=3.9cm]{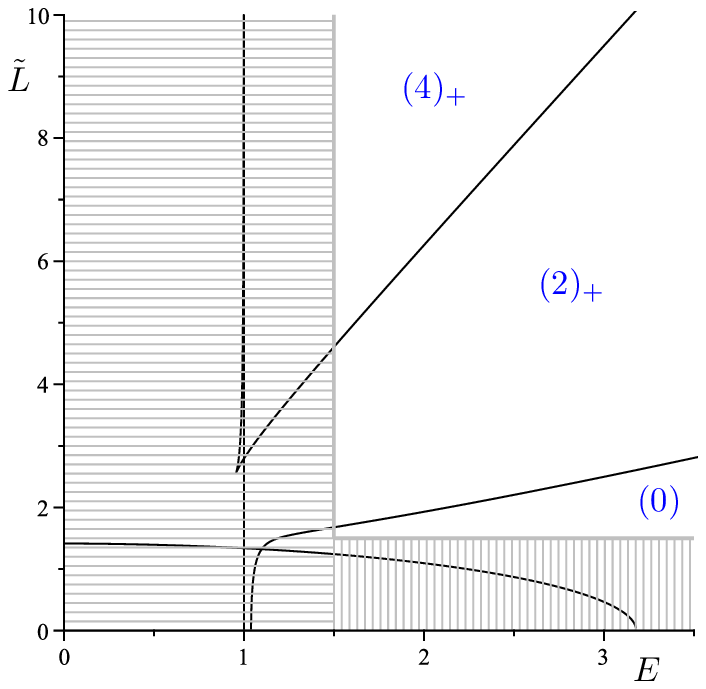}}
\subfigure[][$n=0.5, k=1.0$]{\label{nutn05k1}\includegraphics[width=3.9cm]{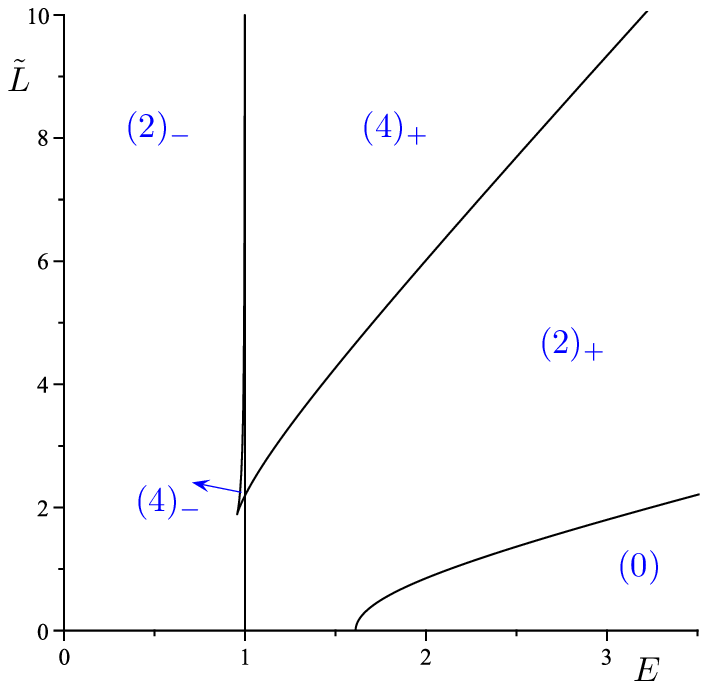}}
\subfigure[][$n=0.5, k=4.55$]{\label{nutn05k455}\includegraphics[width=3.9cm]{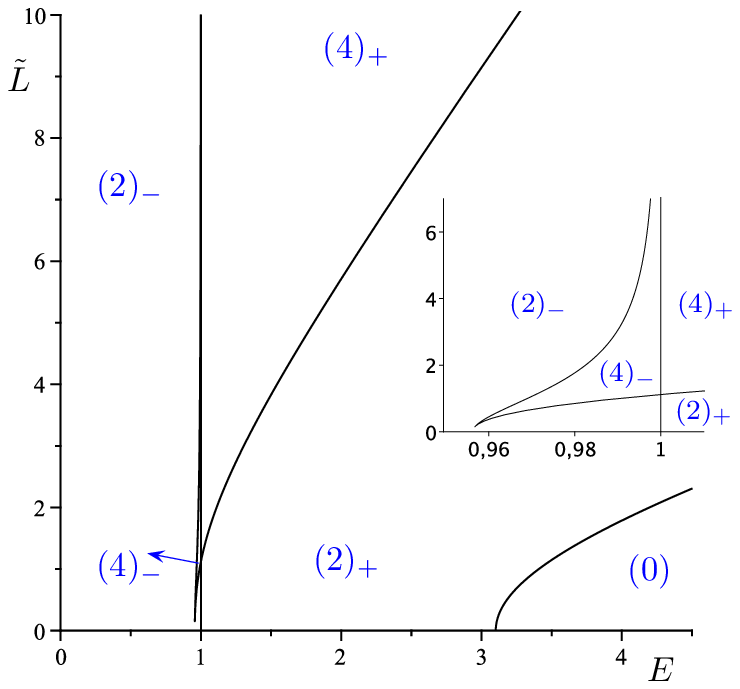}}
\subfigure[][$n=0.5, k=5.55$]{\label{nutn05k555}\includegraphics[width=3.9cm]{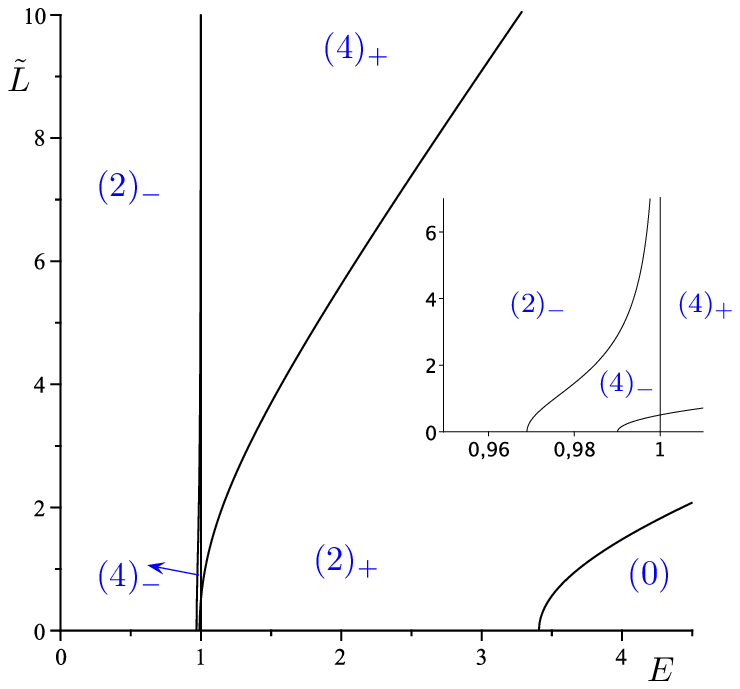}}
\subfigure[][$n=0.5, k=15.0$]{\label{nutn05k15}\includegraphics[width=3.9cm]{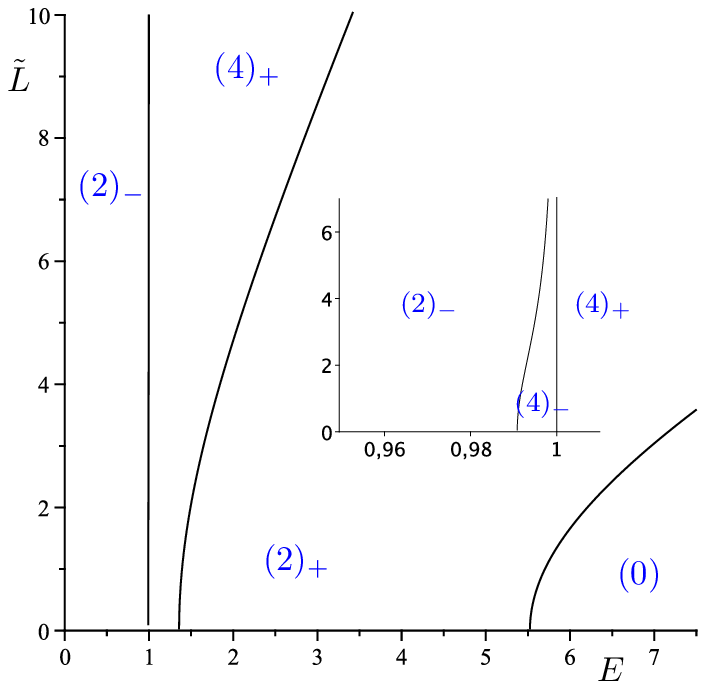}}
\end{center}
\caption{Parametric $\tilde{L}$-$E$ digrams showing the location of the 
regions $(0)$--$(4)$, which reflect the number of zeros of the polynomial $R$ 
in Eq.~\eqref{R_polynomial}. 
Each region contains a set of orbits peculiar to it which are described 
in Table~\ref{TypesOfOrbits1} and in the text (Section~\ref{radmotion}). 
Here the influence of the independent variation of $n$ (fig.\ref{nutn005k0}--\subref{nut5k0}), and $k$ (fig.\ref{nutn05k-2}--\subref{nutn05k15}) on the zeros of $R$ is presented. The dashed region denotes the prohibited values of $E$ and $\tilde L$ following from the inequalities~\eqref{theta-cond-lamu}. Here $C=0$. \label{nut_LE-diagrams}}
\end{figure}

The right hand side of the differential equation~\eqref{eq-r-theta:1} has the form $R=\sum^4_{i=0}{b_i \tr^i}$ with the coefficients 
\begin{eqnarray}
b_4 & = & {E}^2 - \delta  \\ 
b_3 & = & \delta  \label{Rcoefficients} \\ 
b_2 & = & 2 \tilde{n}^2 {E}^2 - c_2  \\ 
b_1 & = & c_2 + \delta \tilde{n}^2 \\ 
b_0 & = & \tilde{n}^2 \left(\tilde{n}^2 {E}^2 + c_2 + \delta \tilde{n}^2 \right)  \, . 
\end{eqnarray} 
Let us now consider massive particles only, that is $\delta = 1$. 
In order to obtain real values for $\tr$ from \eqref{eq-r-theta:1} 
we have to require $R\geq0$. 
The regions for which $R \geq 0$ are bounded by the zeros of $R$. 
The number of zeros depend on the values of $E$, $\tilde L$, and $\tilde n$. 
For $E^2 - 1 \neq 0$ there are regions with $j$ and $j \pm 2$ real zeros 
whose boundaries are given by $R = 0$ and $R^\prime = 0$. 
This is used for the parameter plots shown in Fig.~\ref{nut_LE-diagrams}. 
One has to additionally take care of the change of the sign of $E^2 - 1$ when $E^2$ crosses $E^2 = 1$. Then the sign of $R(\tilde r)$ for $\tilde r \rightarrow \pm \infty$ changes. Furthermore, the case $E^2 = 1$ requires additional attention: If the line $E^2 = 1$ is contained in a region with 4 or 2 zeros then on this portion of the line we have one zero less, that is only 3 or 1 zeros, respectively. Taking all these features into account we obtain the $L$--$E$ diagrams of Fig.~\ref{nut_LE-diagrams}.

Before classifying all orbits we show that it is not possible to have an orbit completely constrained to the region between the two horizons. If such an orbit would exist then the turning points should lie in this region. These turning points are given by
\begin{equation}
0 = \left(\frac{d\tilde r}{d\gamma} \right)^2 = \left(\tilde r^2 + \tilde n^2\right)^2 \left(E^2 - \tilde\Delta_r \frac{\tilde r^2 + \tilde L^2 + k}{\left(\tilde r^2 + \tilde n^2\right)^2}\right) \, ,
\end{equation}
where $\tilde L^2 + k \geq 0$ as can be inferred from \eqref{theta-cond-lamu}. Then it is clear that for $E^2 = 0$ we have $\tilde\Delta_r = 0$ which gives the horizons $r = r_\pm$ as turning points. This case requires $k \geq \tilde n^2$. If $E^2 > 0$ then also $\tilde\Delta_r > 0$, 
implying that the turning points are in the outer regions, 
that is in $r > r_+$ and $r < r_-$. 
Therefore it is not possible to have geodesic motion completely 
constrained to the region between the two horizons. 
Moreover, having crossed one of the horizons once,
a particle is forced to pass the origin $r=0$ and then cross the other 
horizon as well. 

Furthermore, it is not possible to have a turning point at one of the horizons only. If, e.g., $r = r_-$ is a turning point, then $\Delta_r(r_-) = 0$. Then we necessarily have $E^2 = 0$. In this case also $r = r_+$ is a turning point. Therefore it is not possible to have a turning point at one of the horizons only. If $E^2 = 0$ then the turning points at the horizon are the only turning points. That is, if one of the turning points is on one of the horizons then a second one is lying on the other horizon, and there are no further turning points. 

The expression 
\begin{equation}
V_{\rm eff} := \tilde\Delta_r \frac{\tilde r^2 + \tilde L^2 + k}{\left(\tilde r^2 + \tilde n^2\right)^2}
\end{equation}
may be regarded as an effective potential. Though it does not determine the motion of the particles in the usual sense it determines the turning points through the condition $E^2 - V_{\rm eff} = 0$.

Using the $\tilde L$--$E$ diagrams in Fig.~\ref{nut_LE-diagrams} as well as the above considerations we can give all possible combinations of zeros of $R$ and an interpretation in terms of specific types of orbits which are summarized in Table~\ref{TypesOfOrbits1}. 
Note that at the moment we are not discussing features related to singularities and/or geodesic incompleteness. This discussion will follow in Section \ref{Sec:Singularities}. 
In the present section we are just exploring the types of orbits
that are mathematically possible 
as solutions of the geodesic equations.

The types of orbits related to various parameters are given by:  
\begin{itemize}
\item Region (0): no real zeros. We have TOs only with particles moving from $\pm \infty$ to $\mp \infty$ which we call orbit A.
\item Region (1): one real zero. Here the coefficient of the highest power in $R$, given by $E^2 - 1$ for massive test particles, vanishes for $E^2=1$. On the $E^2=1$ line within the region (2) there is one real zero. The orbit is a CEO with particles coming from $\tilde r = + \infty$. 
\item Region (2): two real zeros. Due to the fact that the coefficient of the $\tr^4$--term is given by $E^2 - 1$ the sign and, thus, the type of orbit changes for $E^2$ larger or smaller than 1. For $E^2 < 1$ only CBOs are possible.
\begin{itemize}
	\item Region $(2)_+$: We obtain EOs and CEOs. 
	\item Region $(2)_-$: Due to the sign change mentioned above the 
types of orbits change leading to one CBO. For $0 < E^2 < 1$ the two turning points are larger than $r_+$ and smaller than $r_-$, respectively. For $E^2 = 0$ the two turning points are lying on the two horizons. 
\end{itemize}
\item Region (3): three real zeros. Again, for $E^2 = 1$ the term of highest power in $R$ vanishes giving three zeros for the $E^2=1$--line contained within the region (4). For $E^2 > 0$ the turning points cannot coincide with the horizons. 
\item Region (4): four real zeros. Here again the types of orbits change 
depending on the sign of the coefficient $E^2-1$ of the leading term in $R$. 
\begin{itemize}
	\item Region $(4)_+$: We obtain two EOs and a CBO. For the CBO the turning points cannot be on the horizons. 
	\item Region $(4)_-$: Here we find BOs and CBOs. The turning points cannot be on the horizons. 
\end{itemize} 
\end{itemize} 

\begin{figure}[th!]
\begin{center}
\subfigure[][$V_{\rm eff}$ for $\tilde{n}=0.5$, ${k}=1.0$, $\tilde{L}=3.0$.]{\label{pot2}\includegraphics[width=0.55\textwidth]{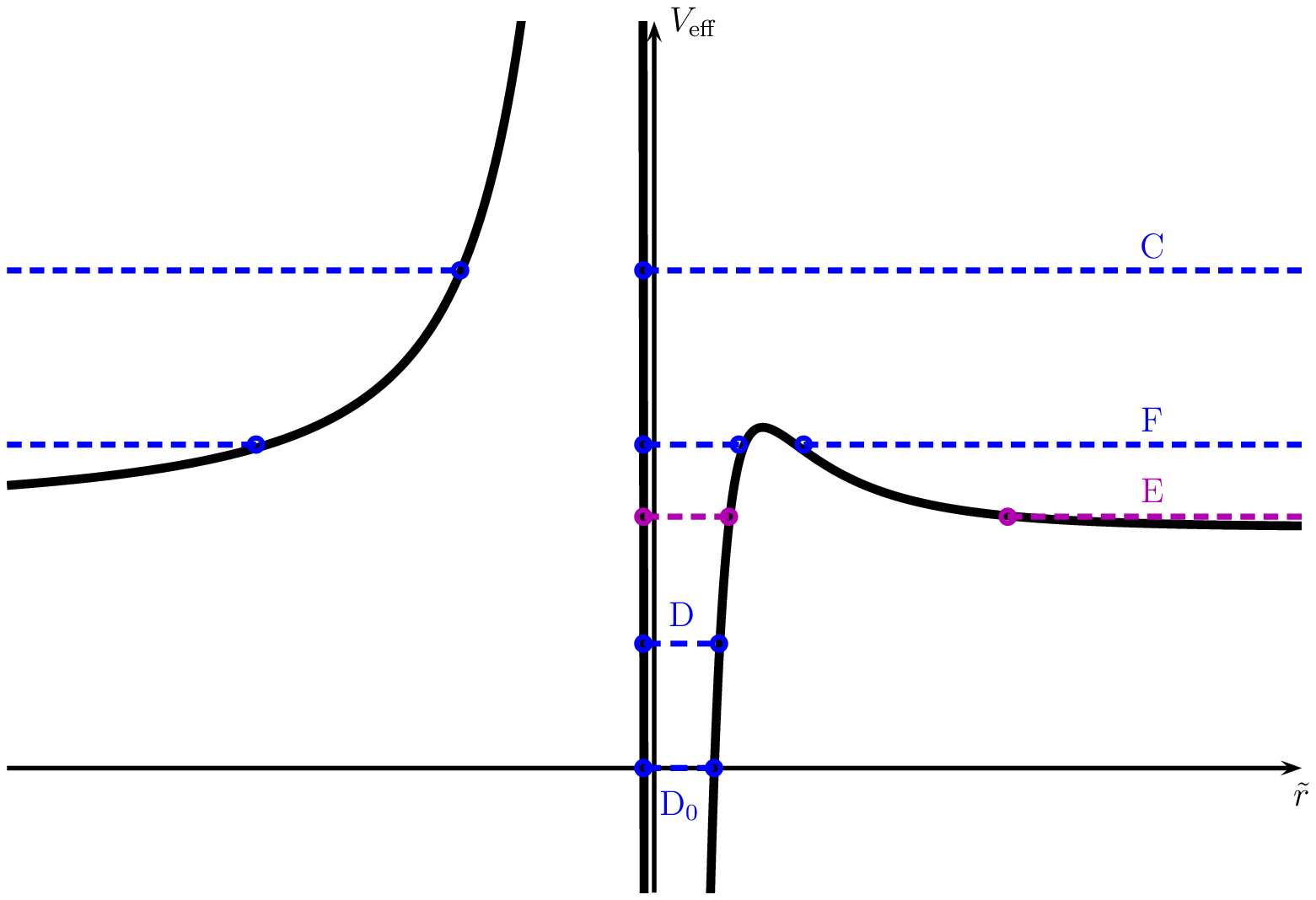}} \qquad
\subfigure[][$V_{\rm eff}$ for $\tilde{n}=0.5$, ${k}=1.0$, $\tilde{L}=2.0$.]{\label{pot1}\includegraphics[width=0.55\textwidth]{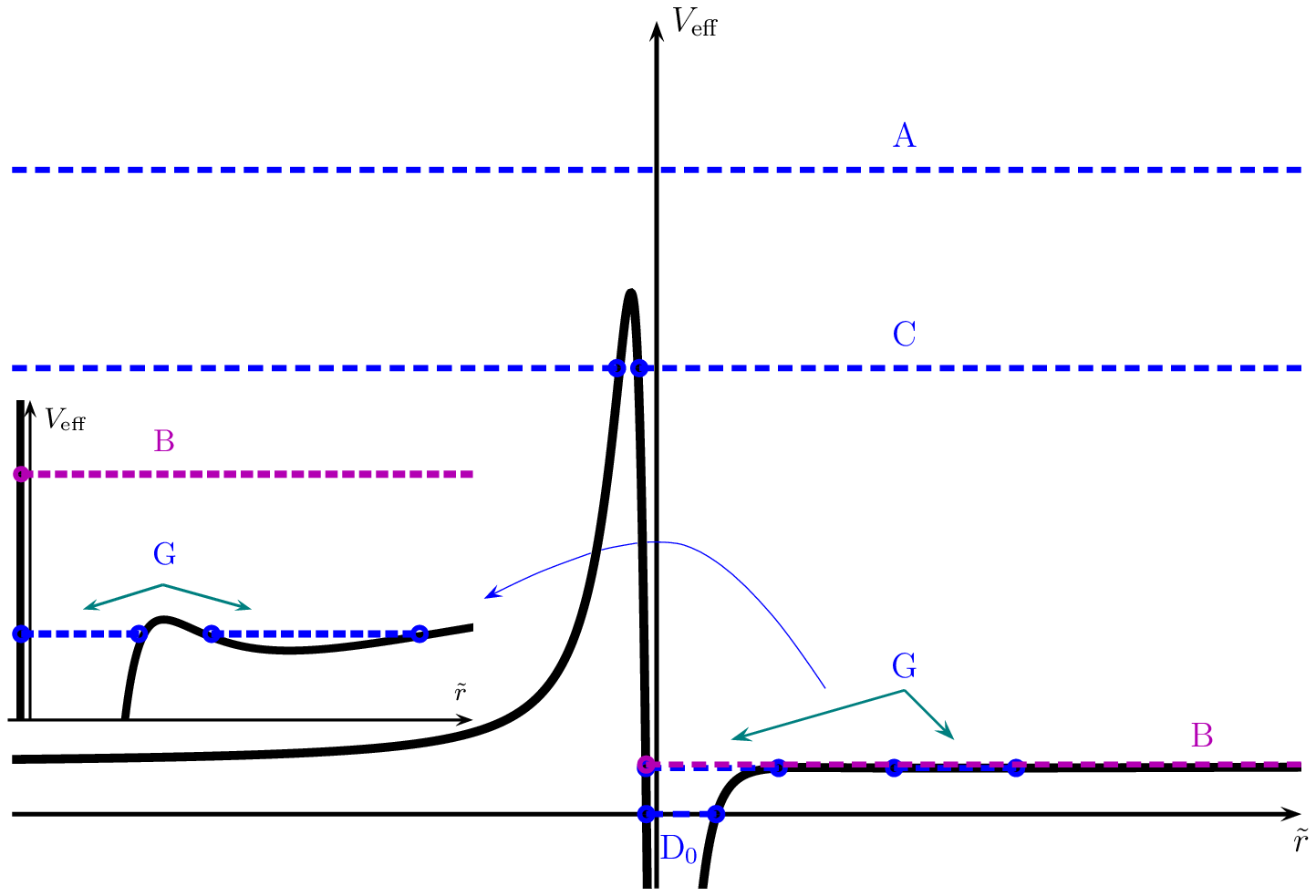}} 
\end{center}
\caption{$V_{\rm eff}$ for orbits of type A, B, C, D, $\rm D_0$, E, F, G 
from Table~\ref{TypesOfOrbits1} showing the position of the orbit types
CBO, CEO, BO, EO, TO. 
The points indicate the turning points of the motion 
and the dashed lines correspond to the values of the energy $E^2$. 
At infinity the effective potential tends to the value
$\lim_{\tr\rightarrow\pm\infty}V_{\rm eff}=1$.
} 
\label{nut_pot} 
\end{figure}

The corresponding effective potentials  
are exhibited in Figure~\ref{nut_pot}.
We note the asymmetry of the possible types of orbits
with respect to the outer regions. In the negative $r$ region
no bound orbits are possible, since the effective potential is
repulsive here.

\begin{table}[t]
\begin{center}
\begin{tabular}{|ccccll|}\hline
type & region & +zeros & --zeros & range of $\tilde r$ & orbit \\ \hline\hline
A & (0) & 0 & 0 & 
\begin{pspicture}(-2,-0.2)(3,0.2)
\psline[linewidth=0.5pt]{->}(-2,0)(3,0)
\psline[linewidth=0.5pt](0,-0.2)(0,0.2)
\psline[linewidth=0.5pt,doubleline=true](1.0,-0.2)(1.,0.2)
\psline[linewidth=0.5pt,doubleline=true](-0.5,-0.2)(-0.5,0.2)

\psline[linewidth=1.2pt](-2,0)(3,0)
\end{pspicture} 
& TO \\  \hline
B & (1) & 0 & 1 & 
\begin{pspicture}(-2,-0.2)(3,0.2)
\psline[linewidth=0.5pt]{->}(-2,0)(3,0)
\psline[linewidth=0.5pt](0,-0.2)(0,0.2)
\psline[linewidth=0.5pt,doubleline=true](1.0,-0.2)(1.0,0.2)
\psline[linewidth=0.5pt,doubleline=true](-0.5,-0.2)(-0.5,0.2)
\psline[linewidth=1.2pt]{*-}(-1.1,0)(3,0)
\end{pspicture} 
 & CEO \\ \hline
C & $(2)_+$ & 0 & 2 & 
   \begin{pspicture}(-2,-0.2)(3,0.2)
\psline[linewidth=0.5pt]{->}(-2,0)(3,0)
\psline[linewidth=0.5pt](0,-0.2)(0,0.2)
\psline[linewidth=0.5pt,doubleline=true](1.0,-0.2)(1.0,0.2)
\psline[linewidth=0.5pt,doubleline=true](-0.5,-0.2)(-0.5,0.2)
\psline[linewidth=1.2pt]{-*}(-2,0)(-1.5,0)
\psline[linewidth=1.2pt]{*-}(-1.0,0)(3,0)
\end{pspicture} 
& EO, CEO \\ \hline
D & $(2)_-$ & 1 & 1 & 
\begin{pspicture}(-2,-0.2)(3,0.2)
\psline[linewidth=0.5pt]{->}(-2,0)(3,0)
\psline[linewidth=0.5pt](0,-0.2)(0,0.2)
\psline[linewidth=0.5pt,doubleline=true](1.0,-0.2)(1.0,0.2)
\psline[linewidth=0.5pt,doubleline=true](-0.5,-0.2)(-0.5,0.2)
\psline[linewidth=1.2pt]{*-*}(-1.0,0)(1.5,0)
\end{pspicture} 
& CBO \\  
${\rm D}_0$ &  &  &  & 
\begin{pspicture}(-2,-0.2)(3,0.2)
\psline[linewidth=0.5pt]{->}(-2,0)(3,0)
\psline[linewidth=0.5pt](0,-0.2)(0,0.2)
\psline[linewidth=0.5pt,doubleline=true](1.0,-0.2)(1.0,0.2)
\psline[linewidth=0.5pt,doubleline=true](-0.5,-0.2)(-0.5,0.2)
\psline[linewidth=1.2pt]{*-*}(-0.5,0)(1,0)
\end{pspicture} 
& ${\rm CBO}_0$ \\  \hline
E & (3) & 2 & 1 & 
\begin{pspicture}(-2,-0.2)(3,0.2)
\psline[linewidth=0.5pt]{->}(-2,0)(3,0)
\psline[linewidth=0.5pt](0,-0.2)(0,0.2)
\psline[linewidth=0.5pt,doubleline=true](1.0,-0.2)(1.0,0.2)
\psline[linewidth=0.5pt,doubleline=true](-0.5,-0.2)(-0.5,0.2)
\psline[linewidth=1.2pt]{*-*}(-1.0,0)(1.5,0)
\psline[linewidth=1.2pt]{*-}(2.0,0)(3,0)
\end{pspicture} 
 & CBO, EO \\ \hline
F & $(4)_+$ & 2 & 2 & 
\begin{pspicture}(-2,-0.2)(3,0.2)
\psline[linewidth=0.5pt]{->}(-2,0)(3,0)
\psline[linewidth=0.5pt](0,-0.2)(0,0.2)
\psline[linewidth=0.5pt,doubleline=true](1.0,-0.2)(1.0,0.2)
\psline[linewidth=0.5pt,doubleline=true](-0.5,-0.2)(-0.5,0.2)
\psline[linewidth=1.2pt]{-*}(-2,0)(-1.5,0)
\psline[linewidth=1.2pt]{*-*}(-1.0,0)(1.5,0)
\psline[linewidth=1.2pt]{*-}(2.0,0)(3,0)
\end{pspicture} 
 & EO, CBO, EO \\ \hline
G & $(4)_-$ & 3 & 1 & 
\begin{pspicture}(-2,-0.2)(3,0.2)
\psline[linewidth=0.5pt]{->}(-2,0)(3,0)
\psline[linewidth=0.5pt](0,-0.2)(0,0.2)
\psline[linewidth=0.5pt,doubleline=true](1.0,-0.2)(1.0,0.2)
\psline[linewidth=0.5pt,doubleline=true](-0.5,-0.2)(-0.5,0.2)
\psline[linewidth=1.2pt]{*-*}(-1.0,0)(1.5,0)
\psline[linewidth=1.2pt]{*-*}(2.0,0)(2.6,0)
\end{pspicture} 
 & CBO, BO  \\ \hline\hline
\end{tabular}
\caption{Types of polynomials and orbits in the Taub--NUT space--time. The thick lines represent the range of orbits. Turning points are shown by thick dots. The regions (1) and (3) are the line $E^2 = 1$ lying in the regions (2) and (4). The indices $+$ and $-$ indicate $E^2 > 1$ and $E^2 < 1$, respectively. The position $\tilde r = 0$ is shown by a vertical line, the horizons are indicated by a vertical double line. In the case ${\rm D}_0$ the turning points lie on the horizons for $E^2=0$.  \label{TypesOfOrbits1}}
\end{center}
\end{table}

\section{Solution of the geodesic equation}

Now we present the analytical solutions of the differential equations ~\eqref{eq-r-theta:1}--\eqref{dtildetdgamma}. 

\subsection{Solution of the $\vartheta$--equation}\label{vartheta-sol}

The solution of Eq.~\eqref{xieom} with $a<0$ and $D>0$ is given by the elementary function
\begin{equation}
\vartheta(\gamma )=\arccos\Bigl(\frac{1}{2a}\left(  \sqrt{D}\sin\left( \sqrt{-a}\gamma - \gamma^\vartheta_{\rm in}   \right) -b  \right) \Bigr) \ ,
\end{equation}
where $\gamma ^\vartheta_{\rm in}=\sqrt{-a}\gamma _{{\rm in}} - \arcsin \left( \frac{2a\xi_{{\rm in}}+b}{\sqrt{b^2-4ac}}\right)$ and $\gamma_{\rm in}$ is the initial value of $\gamma$.

\subsection{Solution of the $\tr$--equation}

For timelike geodesics the polynomial $R$ in~\eqref{R_polynomial} is of fourth order. 
A standard substitution $\tr=\pm\frac{1}{x} + \tr_R$, where $\tr_R$ is a zero of $R$, reduces \eqref{eq-r-theta:1} to a differential equation with a third order polynomial  $\left(\frac{dx}{d\gamma}\right)^2=R_3$, where $R_3=\sum^3_{i=0}{b_ix^i}$. A further substitution $x=\frac{1}{b_3}\left(4y-\frac{b_2}{3}\right)$ transforms that into the standard Weierstra{\ss} form
\begin{equation}
\left(\frac{dy}{d\gamma}\right)^2=4y^3-g_2y-g_3:=P_3(y) \, ,  \label{P3}
\end{equation}
where 
\begin{equation}
g_2=\frac{b_2^2}{12} - \frac{b_1b_3}{4} \, , \qquad  g_3=\frac{b_1b_2b_3}{48} - \frac{b_0b_3^2}{16}-\frac{b_2^3}{216} \ .
\end{equation}
The differential equation \eqref{P3} is of elliptic type and is solved by the Weierstra{\ss} $\wp$--function~\cite{Markush}
\begin{equation}
y(\gamma) = \wp\left(\gamma - \gamma'_{\rm in}; g_2, g_3\right) \ , \label{soly}
\end{equation}
where $\gamma'_{\rm in}=\gamma_{\rm in}+\int^\infty_{y_{\rm in}}{\frac{dy}{\sqrt{4y^3-g_2y-g_3}}}$
with $y_{\rm in}=\pm\frac{b_3}{4}\left(\tr_{\rm in} - \tr_R\right)^{-1} + \frac{b_2}{12}$.
Then the solution of~\eqref{eq-r-theta:1} acquires the form
\begin{equation}
\tr=\pm \frac{b_3}{4 \wp\left(\gamma - \gamma'_{\rm in}; g_2, g_3\right) - \frac{b_2}{3}} + \tr_R \ . \label{solrNUTlight}
\end{equation}

\subsection{Solution of the $\varphi $--equation}

Eq.~\eqref{dvarphidgamma} can be simplified by using \eqref{eq-r-theta:2} 
and by performing the substitution $\xi = \cos\vartheta$ 
\begin{equation}
d\varphi = - \frac{d\xi}{\sqrt{\Theta_\xi}} \frac{\Lp}{1 - \xi^2} + \frac{\xi d\xi}{\sqrt{\Theta_\xi}} \frac{2 \tilde{n} {E}}{1 - \xi^2} \label{phi_new} \, ,
\end{equation}
where $\Theta_\xi$ is given in \eqref{xieom}. This equation can be easily integrated and the solution for $a < 0$ and $D > 0$ is given by
\begin{equation}
\varphi(\gamma) = \half \Bigl(I_- - I_+\Bigr)\Bigl|^{\xi(\gamma)}_{\xi_{{\rm in}}} + \varphi _{\rm in} \label{sol2phi} \, ,
\end{equation}
where
\begin{eqnarray}
I_{\pm} & = & \frac{\Lp\pm2{E} \tilde{n}}{|\Lp\pm2{E} \tilde{n}|} \arctan\frac{1 - u B_\mp}{\sqrt{1-u^2}\sqrt{B_\mp^2-1}} \label{IpmEq}\\
u & = & \frac{2 a \xi +b}{\sqrt{D}} \\ 
B_\pm & = & \frac{b\pm2a}{\sqrt{D}} \, .
\end{eqnarray}
It can be shown that 
\begin{equation}
B_\pm^2-1 = \frac{4(-a)}{D}(2{E} \tilde{n} \mp \Lp)^2>0 \, .
\end{equation}

For the special case ${c_2}=\Lp{^2}$ and $\Lp = 2 {E} \tilde{n}$ the solution reduces to the simple form
\begin{equation}
\varphi (\gamma ) =  \half \arctan\frac{\sqrt{2}(1 - 3 \xi)}{4\sqrt{-\xi^2+\xi}}\Biggl|^{\xi(\gamma)}_{\xi_{{\rm in}}} + \varphi_{\rm in}   \label{sol2phi2_special} \ .
\end{equation}

\subsection{Solution of the $\tilde{t}$--equation}

Eq.~\eqref{dtildetdgamma} consists of $\tr$ and $\vartheta$ parts:
\begin{eqnarray}
\tilde{t}-\tilde{t}_{\rm in}&=& {E} \int^{\tr(\gamma)}_{\tr_{\rm in}} \frac{\trh^4}{\tDr} \frac{d\tr}{\sqrt{R}} + 2 \tilde{n} \int^{\vartheta(\gamma)}_{\vartheta_{\rm in}} \left( \cos\vartheta + C \right) \frac{\Lp - 2 \tilde{n} {E} \cos\vartheta }{\sin^2\vartheta} \frac{  d\vartheta}{\sqrt{\Theta}} =: I_{\tr}(\gamma) + I_{\vartheta}(\gamma) \ . \label{tint_NUT}
\end{eqnarray}
Again with the substitution $\xi=\cos\vartheta$ the integral $I_{\vartheta}(\vartheta)$ can be easily solved:
\begin{equation}
I_{\vartheta}(\gamma)=\tilde{n}\Bigl( I_+ + I_- \Bigr)\Bigl|^{\xi(\gamma)}_{\xi_{\rm in}} + C \tilde{n}\Bigl( I_- - I_+ \Bigr)\Bigl|^{\xi(\gamma)}_{\xi_{\rm in}} 
+ \frac{4\tilde{n}^2{E}}{\sqrt{-a}}\arcsin{ \frac{2a\xi+b}{\sqrt{D}} } \Bigl|^{\xi(\gamma)}_{\xi_{\rm in}} \label{Ivartheta}
\end{equation} 
with $I_{\pm}$ as in ~\eqref{IpmEq}.

We now consider $I_{\tr}$. 
The substitution $\tr=\pm \frac{b_3}{4y-\frac{b_2}{3}} + \tr_R$ 
reexpresses $I_{\tr}$ in terms of $y$:
\begin{equation}
I_{\tr}(\gamma)= E \int^y_{y_{\rm in}} \mp \frac{dy}{\sqrt{P_3(y)}} \frac{\Biggl\{  \left(  \tr_{\rm R}\left(4y-\frac{b_2}{3}\right)\pm b_3  \right)^2 + n^2\left(4y-\frac{b_2}{3}\right)^2  \Biggr\}^2}{\left(4y-\frac{b_2}{3}\right)^2\Delta_y} \ , \label{IryNUT}
\end{equation}
where $\Delta_y=\tDr(\tr_R)\left(4y-\frac{b_2}{3}\right)^2\pm\left(2 \tr_R -1 \right)b_3\left(4y - \frac{b_2}{3}\right) + b_3^2 = 16 \tDr(\tr_R) (y-p_1)(y-p_2)$. Here, $p_1$ and $p_2$ are two zeros of $\Delta_y$. 

We next apply a partial fractions decomposition upon Eq.~\eqref{IryNUT}
\begin{equation}
I_{\tr}(\gamma)= E \int^\gamma_{\gamma_{\rm in}} \mp \frac{dy}{\sqrt{P_3(y)}} \Biggl( K_0 + \sum^3_{j=1}\frac{K_j}{y-p_j} + \frac{K_4}{\left(y-p_3\right)^2} \Biggr) \ , \label{IryNUT2}
\end{equation}
where $p_3=\frac{b_2}{12}$ and $K_i$, $i=0,..,4$, 
are constants which arise from the partial fractions decomposition. These depend on the parameters of the metric 
and the test particle and on $\tr_{\rm R}$. 
After the substitution $y=\wp(v)$ with 
$\wp^\prime(v)=(-1)^\rho\sqrt{4 \wp^3(v)-g_2\wp(v)-g_3}$, 
where $\rho$ is either $0$ or $1$ depending 
on the sign of $\wp^\prime(v)$ in the considered interval 
and on the branch of the square root, 
Eq.~\eqref{IryNUT2} simplifies to
\begin{equation}
I_{\tr}(\gamma)= \mp E \int^v_{v_{\rm in}} (-1)^\rho \Biggl( K_0 + \sum^3_{j=1}\frac{K_j}{\wp(v)-p_j} + \frac{K_4}{\left(\wp(v)-p_3\right)^2} \Biggr)dv \ . \label{IryNUT3}
\end{equation}  
Here $v=v(\gamma)=\gamma-\gamma^\prime_{\rm in}$ and $v_{\rm in}=v(\gamma_{\rm in})$. 

The final solution takes the form (details can be found in the appendices \ref{elldiffIII} and \ref{elldiff2})
\begin{eqnarray}
I_{\tr}(\gamma) & = & \mp (-1)^\rho E \Biggl\{\left(K_0 + A_2K_4\right)(v-v_{\rm in}) +  \sum^2_{i=1}\Biggl[  \sum^3_{j=1} \frac{K_j}{\wp^\prime(v_{ji})}\Biggl( \zeta(v_{ji})(v-v_{\rm in}) + \log\frac{\sigma(v-v_{ji})}{\sigma(v_{\rm in}-v_{ji})}  \Biggr)  \nonumber \\
&& - \frac{K_4}{\left(\wp^\prime(v_{3i})\right)^2}\Biggl( \zeta(v-v_{3i}) - \zeta(v_{\rm in}-v_{3i}) + \frac{\wp^{\prime\prime}(v_{3i})}{\wp^\prime(v_{3i})}
 \log\frac{\sigma(v-v_{3i})}{\sigma(v_{\rm in}-v_{3i})} \Biggr) \Biggr] \Biggr\}    \ , \label{IryNUT4}
\end{eqnarray} 
where $v_{ji}$ are the poles of the functions $(\wp(v)-p_j)^{-1}$ and $(\wp(v)-p_3)^{-2}$ in~\eqref{IryNUT2} such that $\wp(v_{j1})=p_j=\wp(v_{j2})$ since $\wp(v)$ is an even elliptic function of order two which assumes every value in the fundamental parallelogram with multiplicity two. Here $\zeta(v)$ is the Weierstra{\ss} zeta function and $\sigma(v)$ is the Weierstra{\ss} sigma function~\cite{Markush}; $\displaystyle{A_2=-\sum^{2}_{i=1}\left(  \frac{\wp(v_{3i})}{\left(\wp^\prime(v_{3i})\right)^2} + \frac{\wp^{\prime\prime}(v_{3i})\zeta(v_{3i})}{\left(\wp^\prime(v_{3i})\right)^3}  \right)}$ is a constant. 

\subsection{The complete orbits}

\begin{figure}[t]
\begin{center}
\subfigure[][3d plot. The sphere shows the horizon $r_+$.]{\label{TOrp}\includegraphics[width=0.34\textwidth]{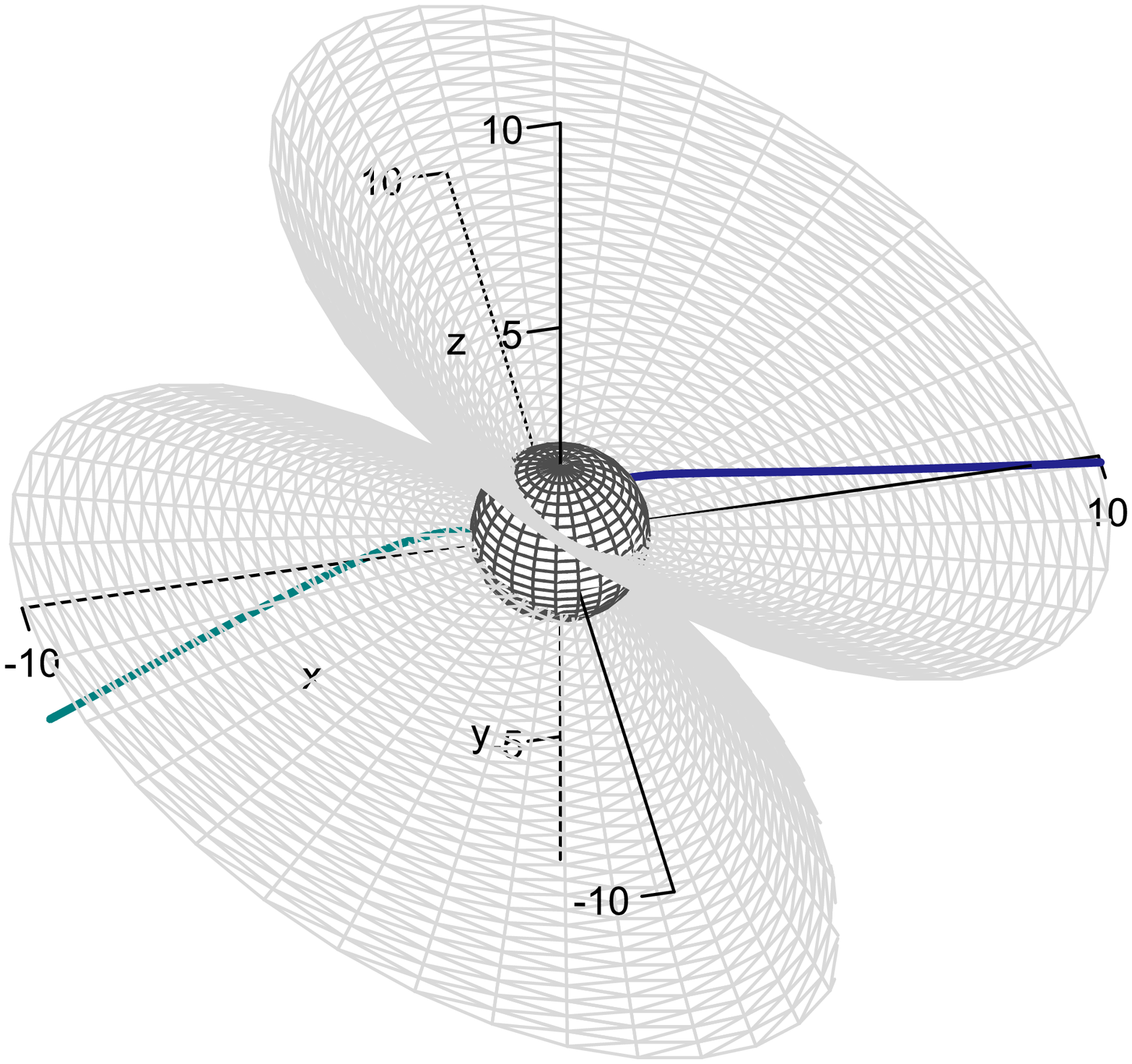}}\qquad
\subfigure[][Projection onto the $x-z$--plane. The two circles denote $r_+$ and $r_-$.]{\label{TOrm}\includegraphics[width=0.34\textwidth]{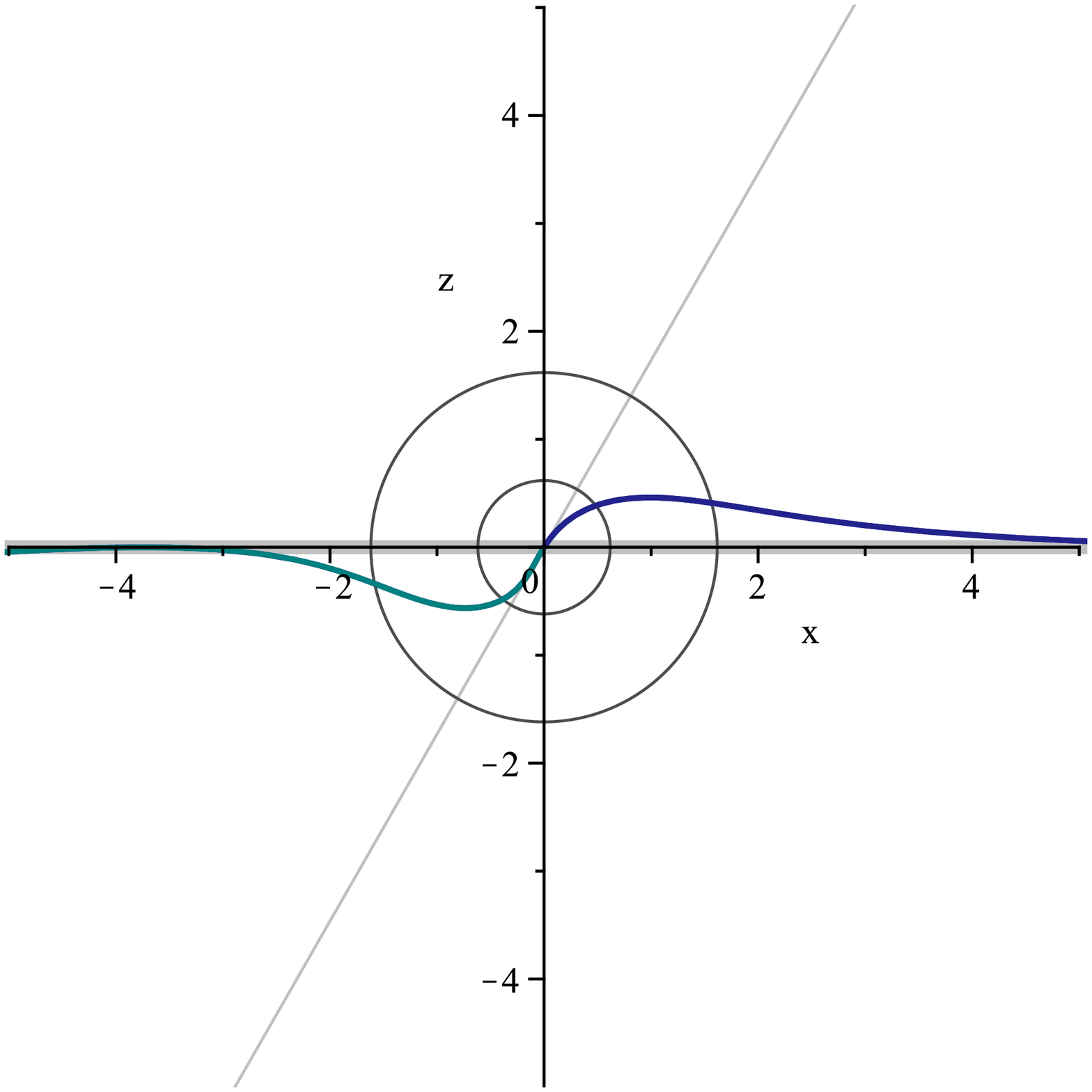}}
\end{center}
\caption{TO with parameters $\tilde n = 1$, $k = 1$, $\tilde L = 2$, $E^2 = 3$.
The figures combine both regions $r > 0$ (blue) and $r < 0$ (sea green). \label{NUTGeodesicsTO}}
\end{figure}

\begin{figure}[t]
\begin{center}
\subfigure[][3d plot. The sphere shows the horizon $r_+$.]{\label{mu093-bound}\includegraphics[width=0.34\textwidth]{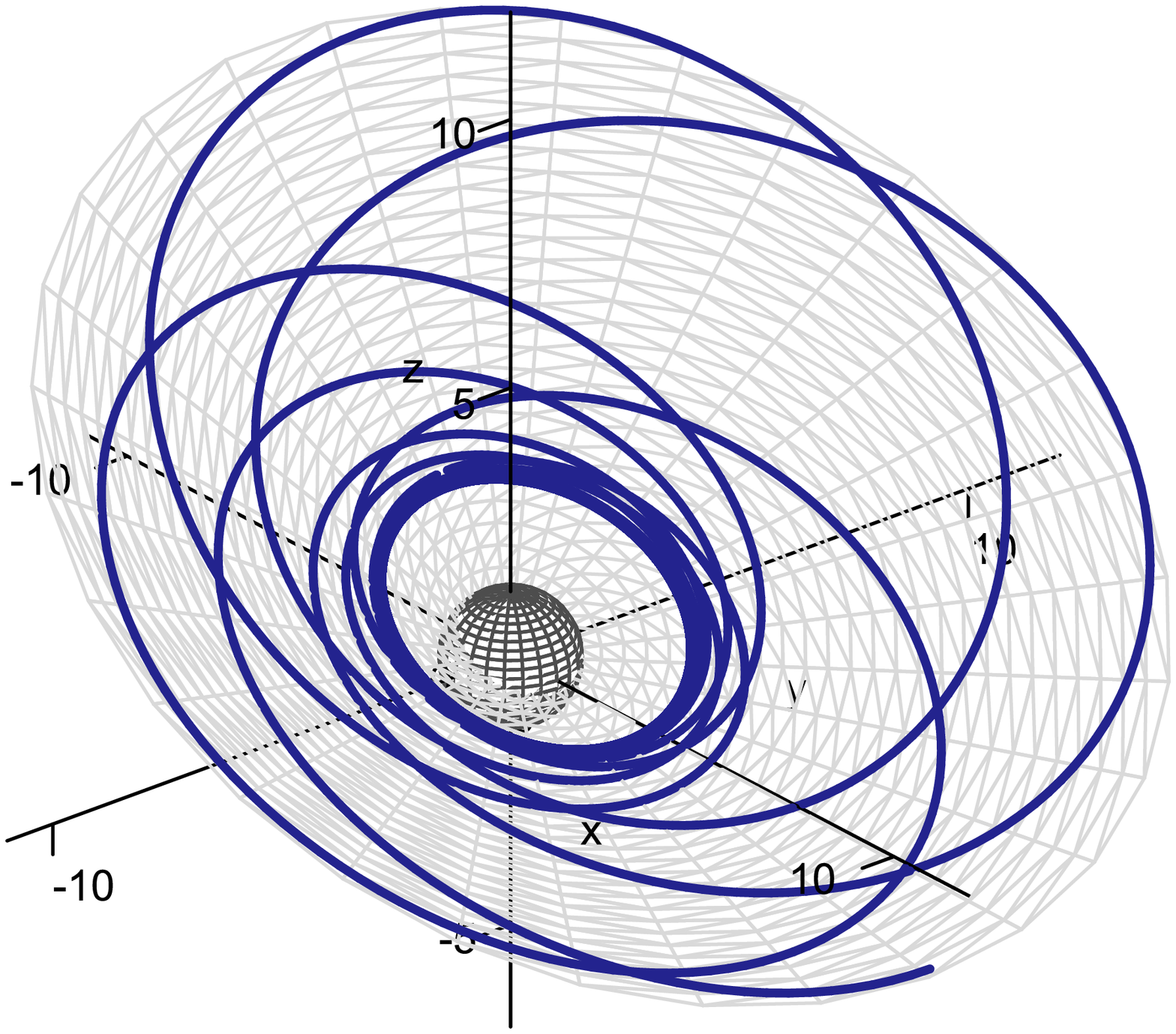}}\qquad
\subfigure[][Projection onto the $x-z$--plane. The two circles denote $r_+$ and $r_-$.]{\label{mu093-boundXY}\includegraphics[width=0.34\textwidth]{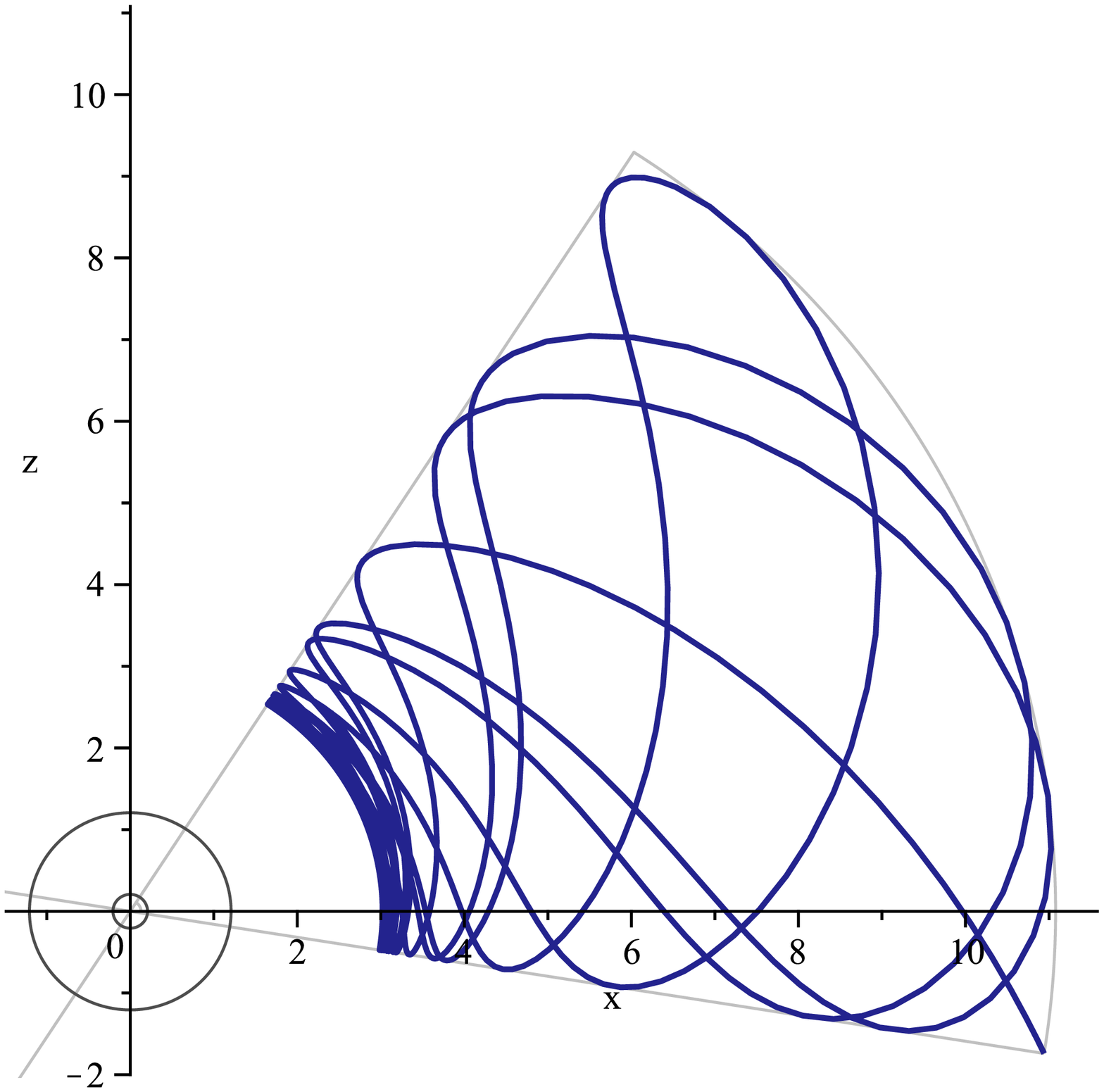}}
\end{center}
\caption{BO with parameters $\tilde{n}=0.5$, ${k}=1.0$, $\tilde{L}=2.0$, ${E}^2=0.94084$. The sphere denotes the horizon at $r_+$. \label{NUTGeodesicsPlotBO}}
\end{figure}

\begin{figure}[t]
\begin{center}
\subfigure[][3d plot. The sphere shows the horizon $r_+$.]{\label{mu093-term}\includegraphics[width=0.34\textwidth]{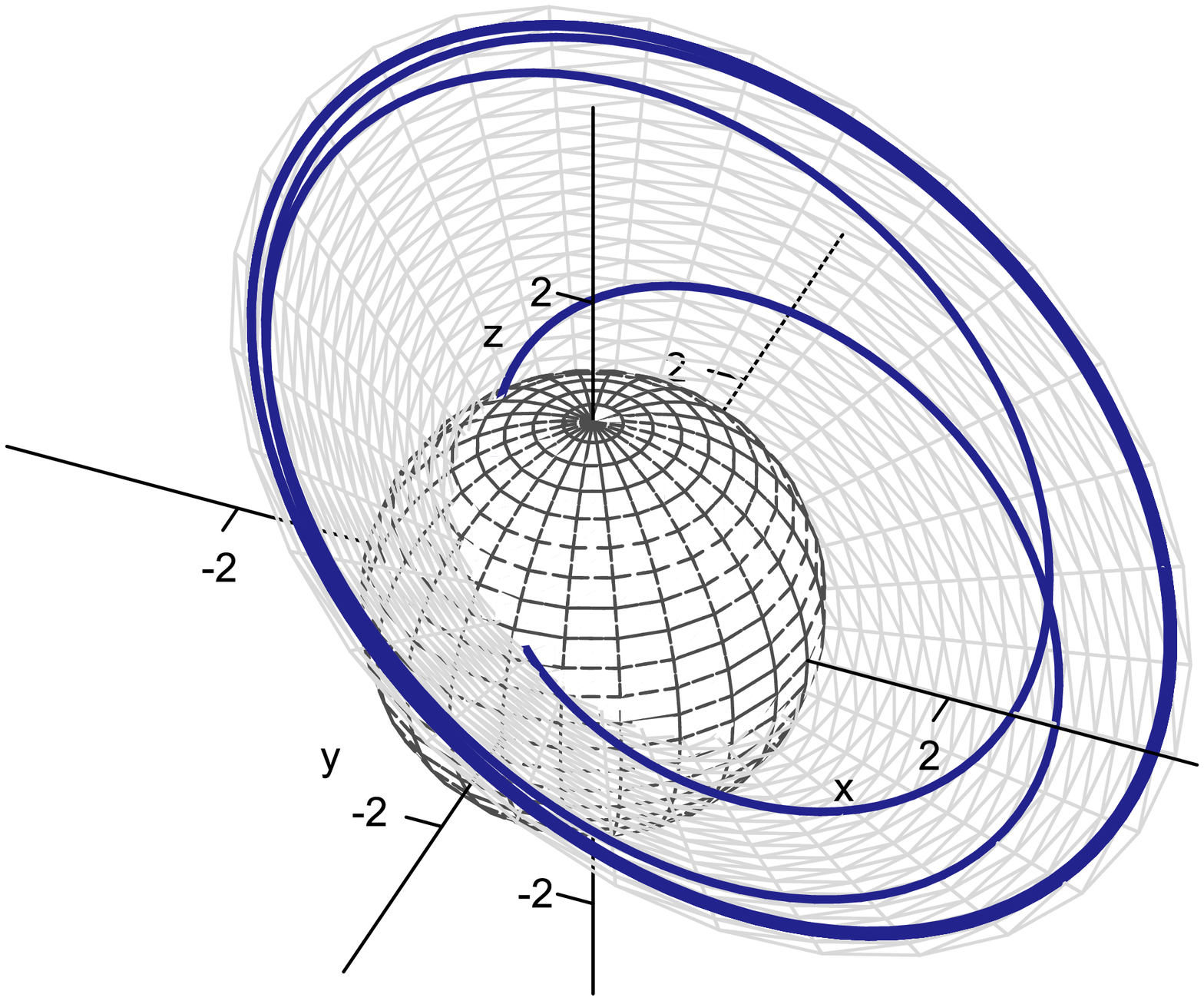}}\qquad
\subfigure[][Projection onto the $x-z$--plane. The two circles denote $r_+$ and $r_-$.]{\label{mu093-termXY}\includegraphics[width=0.34\textwidth]{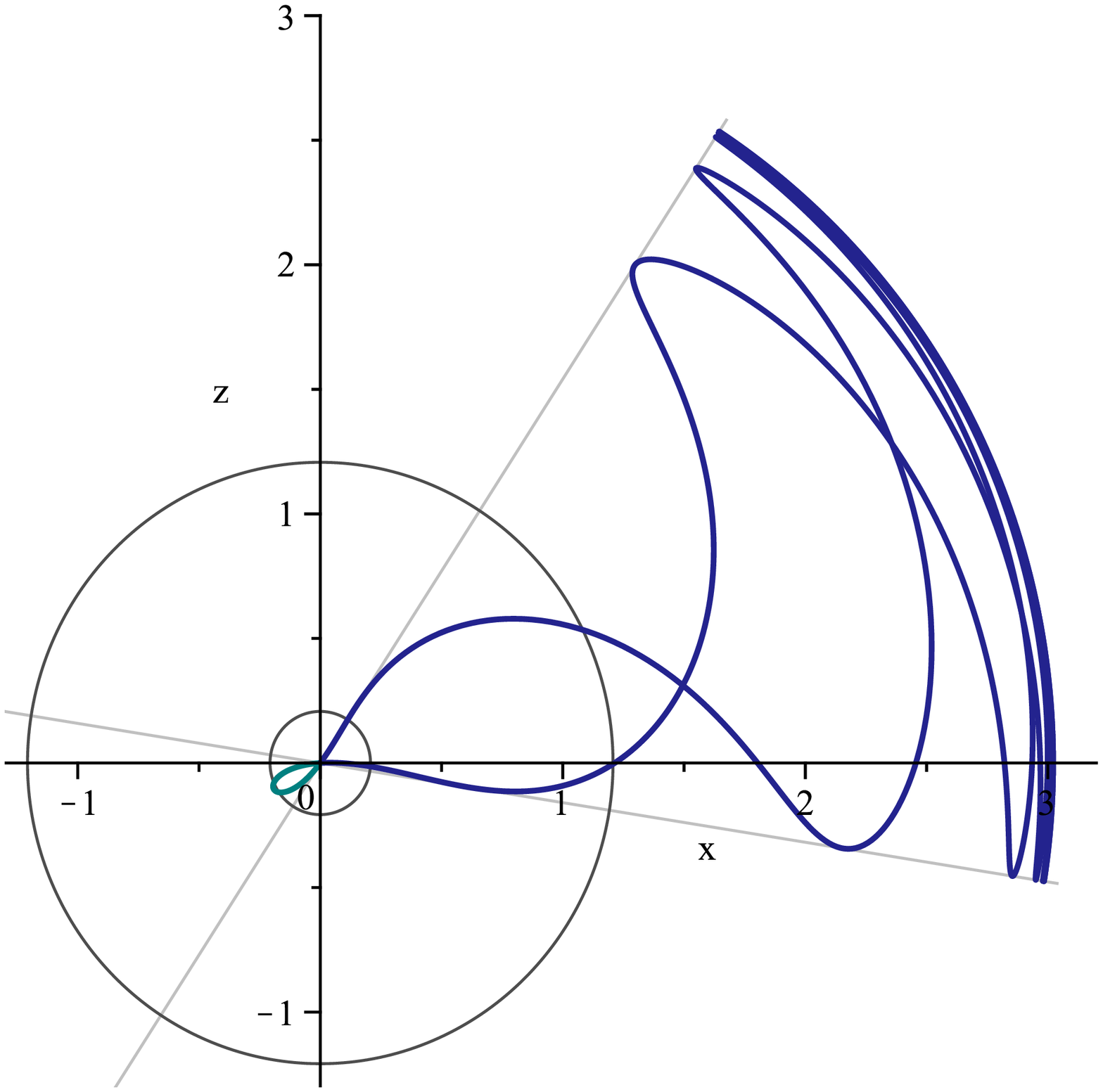}} 
\end{center}
\caption{CBO with parameters $\tilde{n}=0.5$, ${k}=1.0$, $\tilde{L}=2.0$, ${E}^2=0.94084$. In the 3d--plot the orbit for positive $r$ (blue) is shown. The transition to negative $r$ (sea green) can be seen in the projection onto the $x-z$--plane. \label{NUTGeodesicsPlotCBO}} 
\end{figure}

\begin{figure}[t]
\begin{center}
\subfigure[][EO with parameters $\tilde{n}=0.5$, ${k}=1.0$, $\tilde{L}=3.0$, ${E}^2=1.369$.]{\label{mu1369-escape}\includegraphics[width=0.34\textwidth]{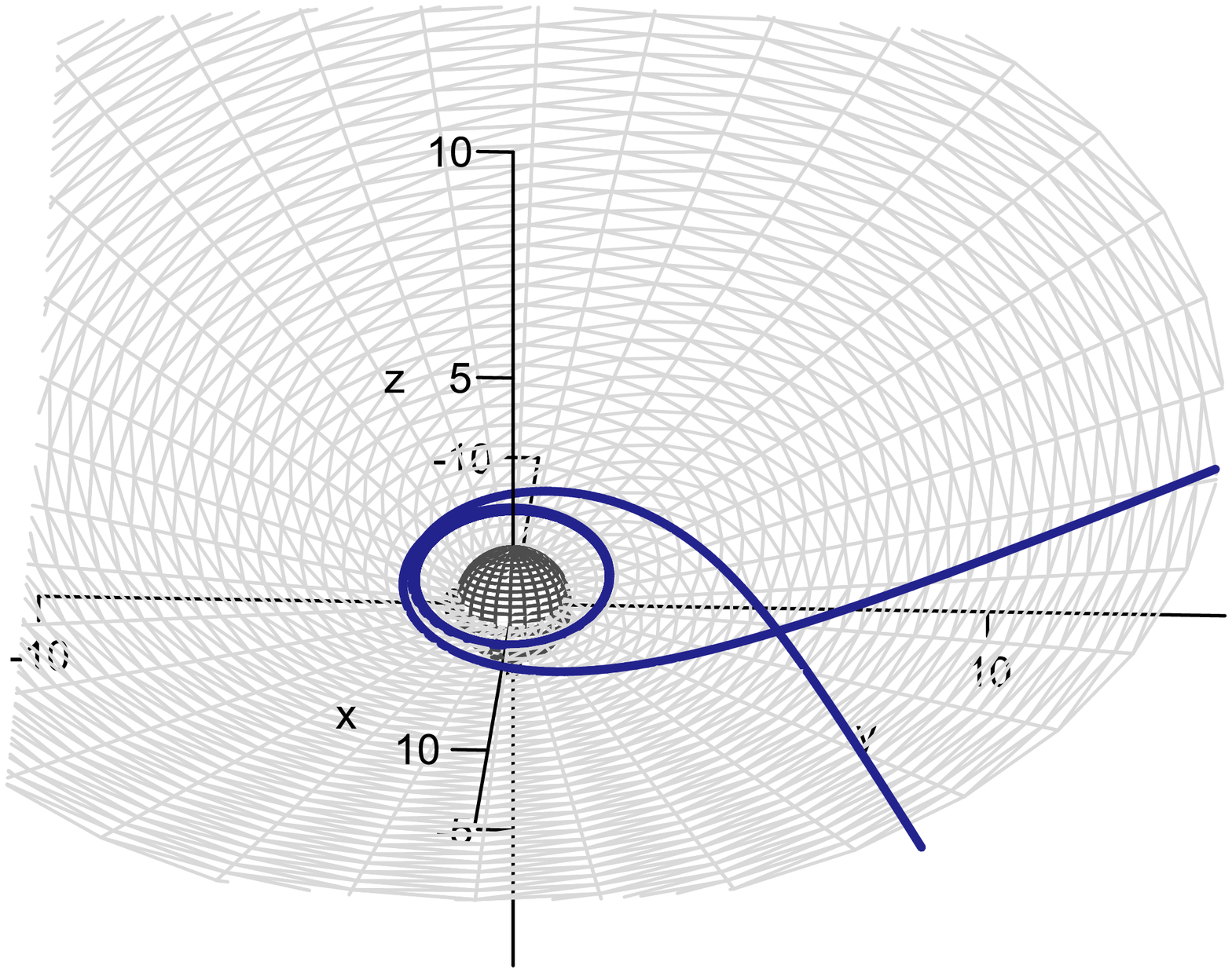}} \qquad
\subfigure[][Projection onto the $x-z$--plane. The two circles denote $r_+$ and $r_-$.]{\label{mu1369-escapeXZ}\includegraphics[width=0.32\textwidth]{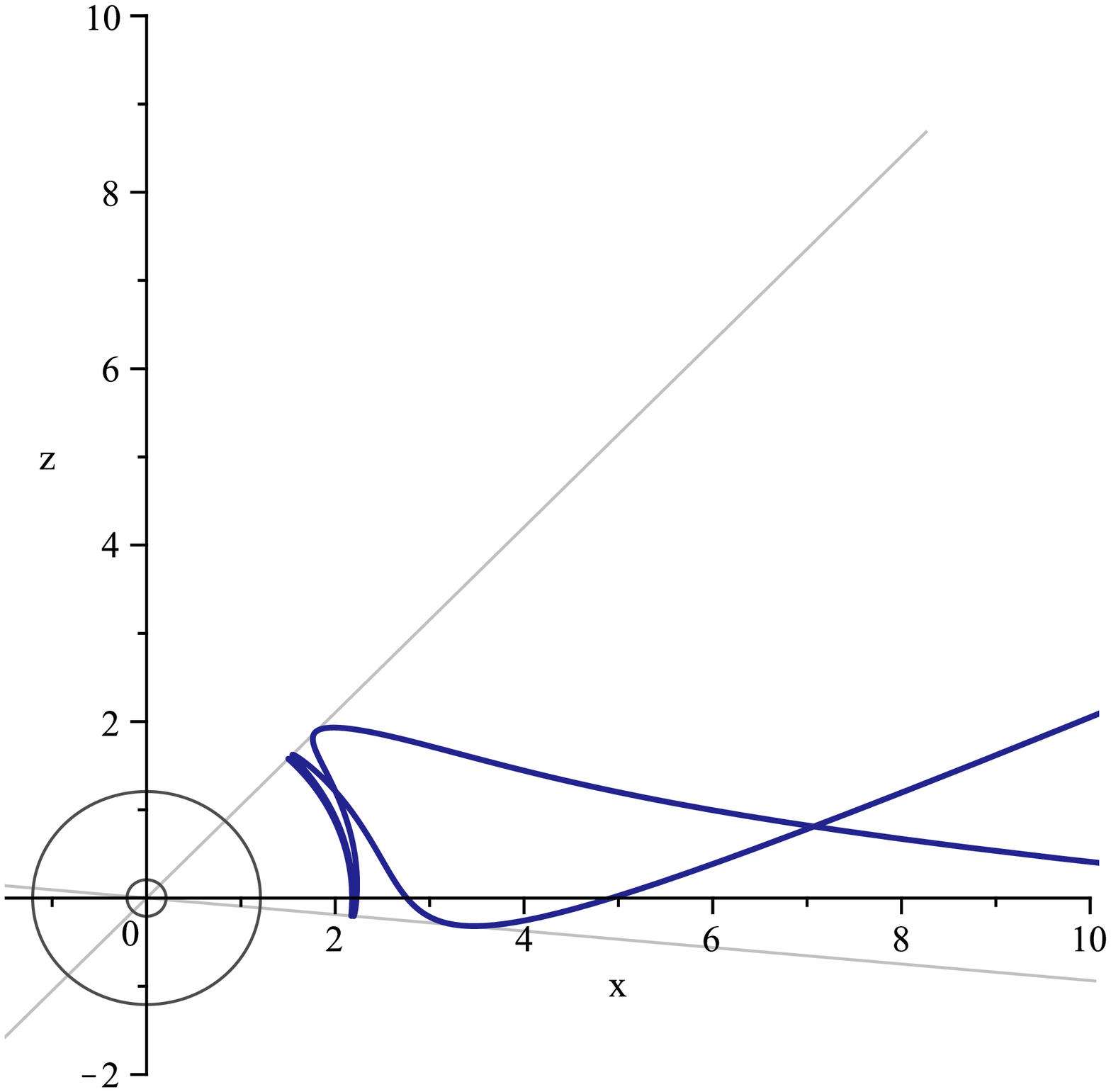}}
\end{center}
\caption{An example for an EO. \label{escape_NUT}}
\end{figure}

\begin{figure}[t]
\begin{center}
\subfigure[][CEO with parameters $n=0.5$, ${k}=-0.5$, $\tilde{L}=2.0$, $E=1.0000001$.]{\label{neg_k}\includegraphics[width=0.34\textwidth]{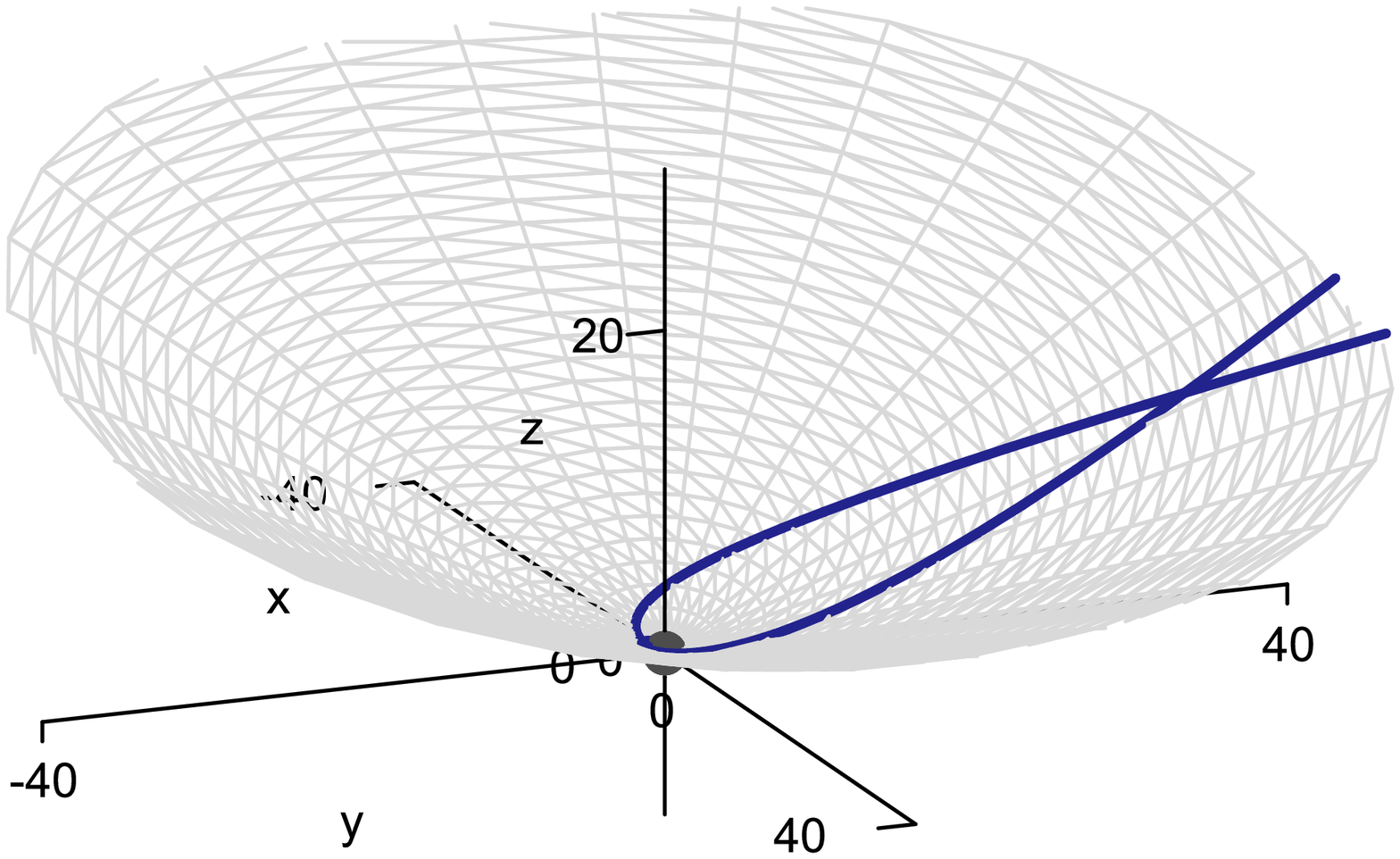}} \qquad
\subfigure[][Projection onto the $x-z$--plane. The two circles denote $r_+$ and $r_-$.]{\label{neg_k_XZ}\includegraphics[width=0.3\textwidth]{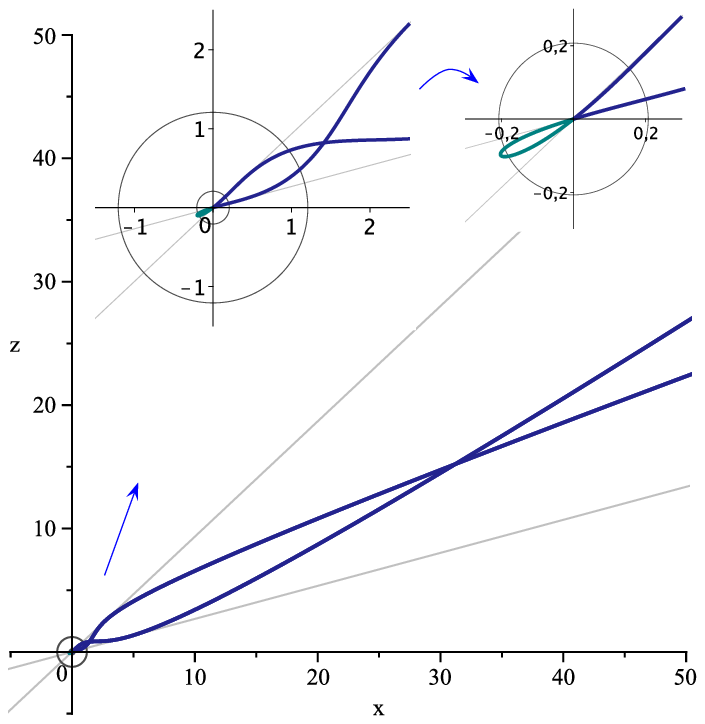}}
\end{center}
\caption{An example for a CEO. The figures combine both regions $r > 0$ (blue) and $r < 0$ (sea green). \label{CEO_NUT}}
\end{figure}

\begin{figure}[t]
\begin{center}
\subfigure[][Degenerate CBO with parameters $n=0.5$, ${k}=1.0$, 
$\tilde{L}=3.0$, $E=0$. The turning points coincide with the horizons, 
where $r_- < 0$ is the inner circle. The motion takes place
in a plane. The figure combines both regions $r > 0$ (blue) and $r < 0$ (sea green).]
{\label{E=0}\includegraphics[width=0.34\textwidth]{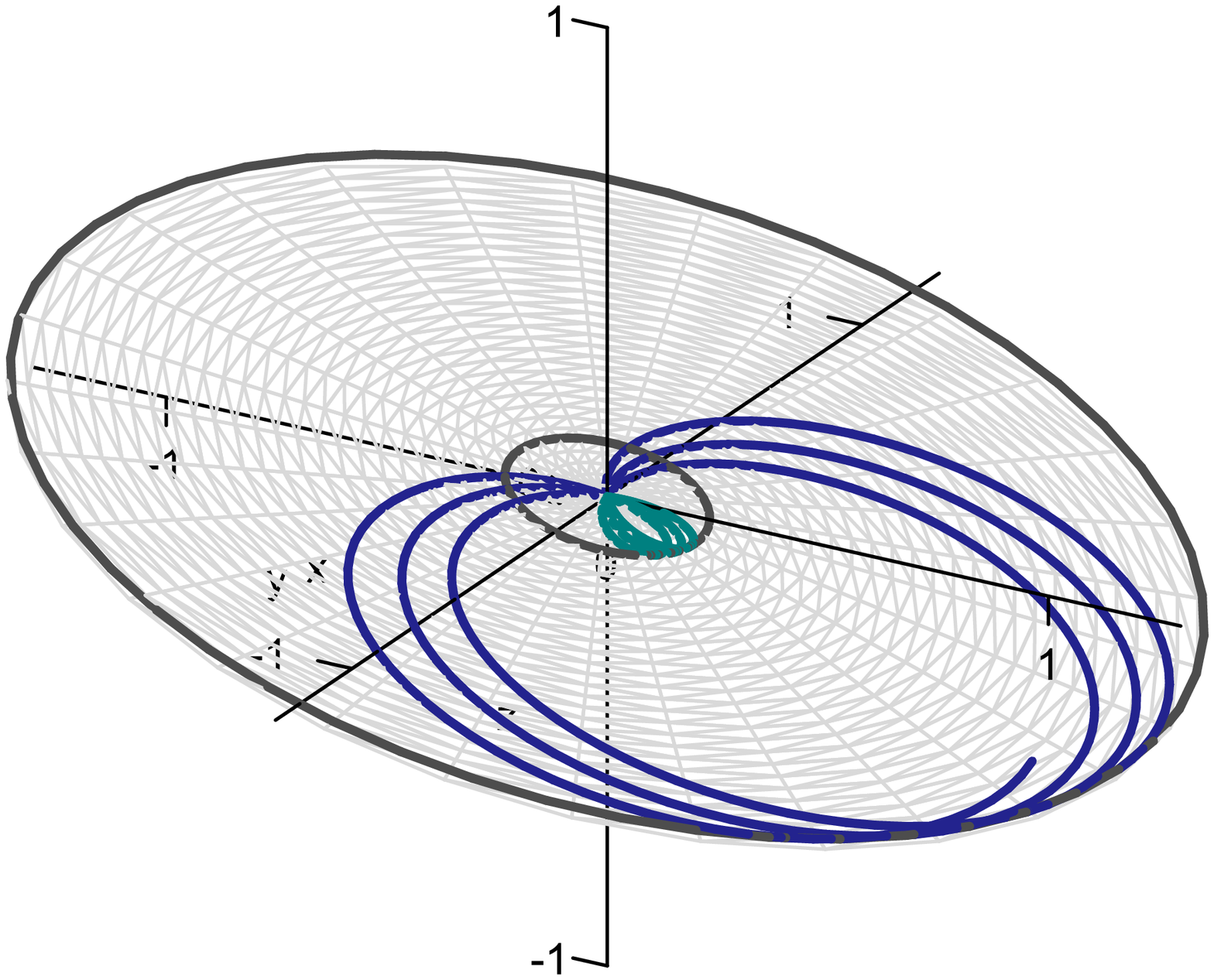}} \qquad 
\subfigure[][CBO: $n=0.5$, ${k}=\tilde{n}^2$, $\tilde{L}=0.9$, 
$E = \tilde{L}/(2 \tilde{n})$. The red point indicates the $\vartheta=0$-axis. 
In the Misner interpretation the axis is regular 
and a test particle can move across $\vartheta=0$; 
alternatively, when the axis is interpreted to represent a physical
singularity, the orbit must end when it encounters this singularity. 
Details are given in Section~\ref{Sec:Singularities}.]
{\label{L=2En}\includegraphics[width=0.39\textwidth]{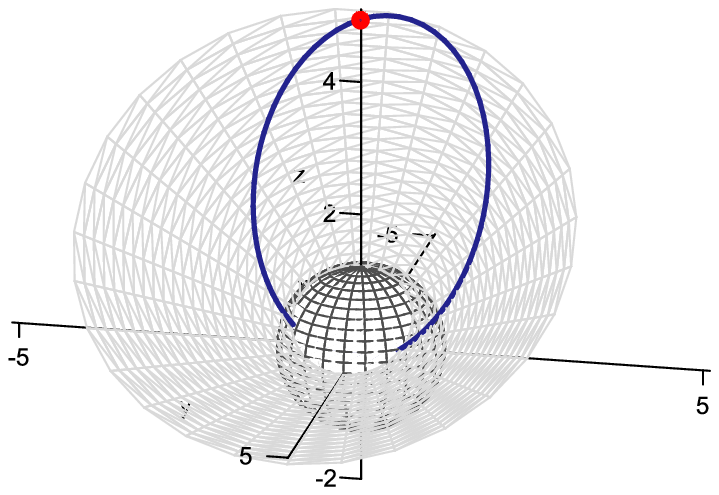}}
\end{center}
\caption{
Two degenerate orbits. A CBO with the horizons as turning points 
and a CBO crossing the $\vartheta=0, \pi$ axis. \label{NUTGeodesicsPlotZaxis}} 
\end{figure}

With these analytical results we have found the full set of orbits 
for massive point particles: 
A TO is shown in Fig.~\ref{NUTGeodesicsTO}, 
a BO in Fig.~\ref{NUTGeodesicsPlotBO}, 
a CBO in Fig.~\ref{NUTGeodesicsPlotCBO}, 
an EO in Fig.~\ref{escape_NUT}, 
and a CEO in Fig.~\ref{CEO_NUT}. 
The degenerate case of a CBO where the turning points are lying 
on the horizons is exhibited in Fig.~\ref{E=0}. A further degenerate case, an orbit which crosses the $\vartheta=0$--axis, is shown in Fig.~\ref{L=2En}. The CBO, CEO and TO contain a positive $r$ and a negative $r$ part. For simplicity and since it does not create confusion, 
these two parts are included in the same plot. In Fig.~\ref{NUTGeodesicsTO}, Fig.~\ref{mu093-termXY}, Fig.~\ref{neg_k_XZ} and Fig.~\ref{E=0} the part of the orbit with negative $r$ is plotted in sea green, whereas the blue color always represents positive $r$. 

The constant $C$ appearing in the general family of Taub--NUT solutions (eq.~\eqref{metrikNUTdeSitter}) is chosen zero in the calculation of the geodesics. In general, it follows from the Hamilton-Jacobi equations~\eqref{eq-r-theta:1}-\eqref{dvarphidgamma} that the constant $C$ can be absorbed in the constant $\Lp$ and does not influence the characteristic features of the geodesics in the Taub-NUT space-times discussed in section~\ref{theta-r-features}. Moreover, as it is shown in the section~\ref{gauge} one can switch between the $C$-dependent metric~\eqref{metrikNUTdeSitter} and Hamilton-Jacobi equations~\eqref{eq-r-theta:1}-\eqref{dtildetdgamma} and the $C$-independent metric~\eqref{metrikNUTdeSitter2} and Hamilton-Jacobi equations~\eqref{eq-r-theta:1_2}-\eqref{dtildetdgamma_2} by introducing a gauge transformed coordinate $t^\prime$ and interchanging the constants of motion $\Lp$ and $\tilde{L}$, $k^\prime$ and $k$.

These orbits are just solutions of the geodesic equation for the spatial coordinates. The complete discussion of these orbits is possible only by including the solutions for the time coordinate and the proper time, which is critical for the issue of geodesic incompleteness and other singularities. This is done in the next section. 

\section{The issue of singularities}\label{Sec:Singularities}

In this section we discuss the nature of the singularities
which are present in the Taub--NUT space--times.
We address these singularities in terms of the analytic solutions 
of the geodesic equation. 
Different interpretations of the Taub--NUT space--time
give rise to different types of singularities.

\begin{enumerate}
\item The family of Taub--NUT space--times as given by the metric~\eqref{metrikNUTdeSitter} possesses a singularity along the whole $\vartheta=0, \pi$ axis
when $|C| \ne 1$
\cite{Manko2005}, which includes our choice $C=0$,
Eq.~\eqref{metrikNUTdeSitter}.
The choice $|C|=1$ leads only to singular semi--axes
(see e.g.~Misner \cite{Misner63}).
\item In the interpretation of Misner \cite{Misner63}
and Misner and Taub \cite{MisnerTaub69} 
one can avoid the singularity present on the axis 
by employing a periodic identification of the time coordinate. 
Hypersurfaces $r=$ const then possess the topology $S^3$.
In this case it is, however,
not possible to eliminate the geodesic incompleteness 
which arises at the horizons, as discussed below.
\item When analytically extending the space--time at the horizons
by gluing further copies of the space--time
following Miller, Kruskal and Godfrey \cite{MiKruGo71},
one has to retain the singularity along the $\vartheta=0, \pi$ axis. 
Therefore one finds the space--times considered by
Bonnor \cite{Bonnor69} and Manko and Ruiz \cite{Manko2005},
where the axial singularity is interpreted
as a physical singularity.
Such a singularity might possibly be avoided, however,
if an appropriate interior solution were included.
\end{enumerate}

Before discussing the singularities of Taub--NUT space--times 
in terms of the behavior of the geodesics, 
we visualize Taub--NUT space-times using Carter--Penrose diagrams. 

\subsection{Singularities in a single copy of Taub--NUT space--time}\label{ssec:incimplTNUT}

\begin{figure}[th!]
\begin{center}
\subfigure[][Taub--NUT space--time I]{\includegraphics[width=0.45\textwidth]{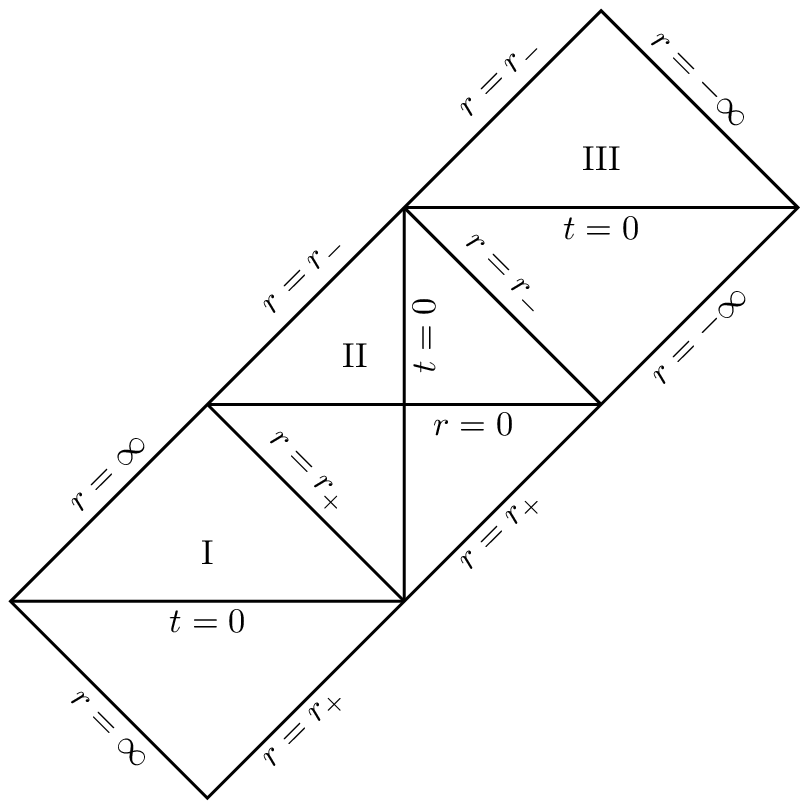}} \qquad
\subfigure[][Taub--NUT space-time II]{\includegraphics[width=0.45\textwidth]{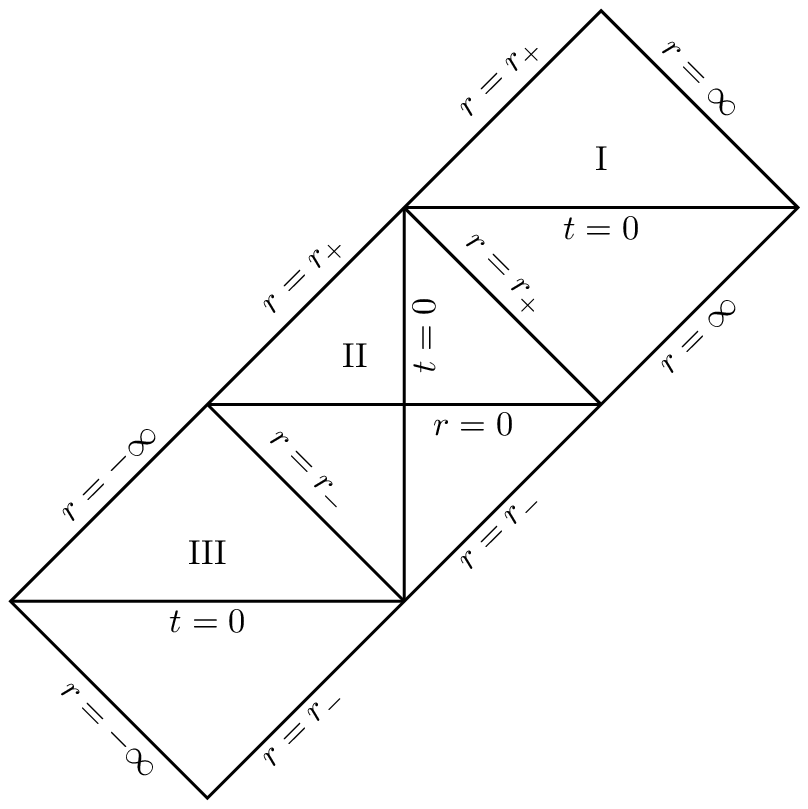}} 
\end{center}
\caption{Two causal versions Taub--NUT space--time in a Carter--Penrose diagram. Here region I is given by $r > r_+$, region II by $r_- < r < r_+$ and region III by $r < r_-$. For the Kruskal--like analytic continuation see Fig.~\ref{Fig:NUTKruskal}. 
\label{Fig:NUTCP} }
\end{figure}

The Carter--Penrose diagram for a single copy of Taub--NUT space--time 
is shown in Fig.~\ref{Fig:NUTCP}. 
It consists of three regions: region I with $\infty > r > r_+ > 0$, 
region II with $r_+ > r > r_-$ and region III with $0 > r_- > r > -\infty$. 
This diagram is helpful in discussing the singularity issues. 
One can consider two versions of it because
the geodesic motion in regions I and III is different. 
(Recall that the potential in region III is repulsive.)
Thus the motion of a particle depends on 
whether it starts in region I or region III.  

\begin{figure}[th!]
\begin{center}
\subfigure[][Taub--NUT space--time I. The CEO (red) terminates at the 
horizon $r_-$, the CBO (blue) starts at the horizon $r_+$ 
and terminates at the horizon $r_-$, 
the orbit ${\rm D}_0$ (green) is a special case of the CBOs.]{\includegraphics[width=0.45\textwidth]{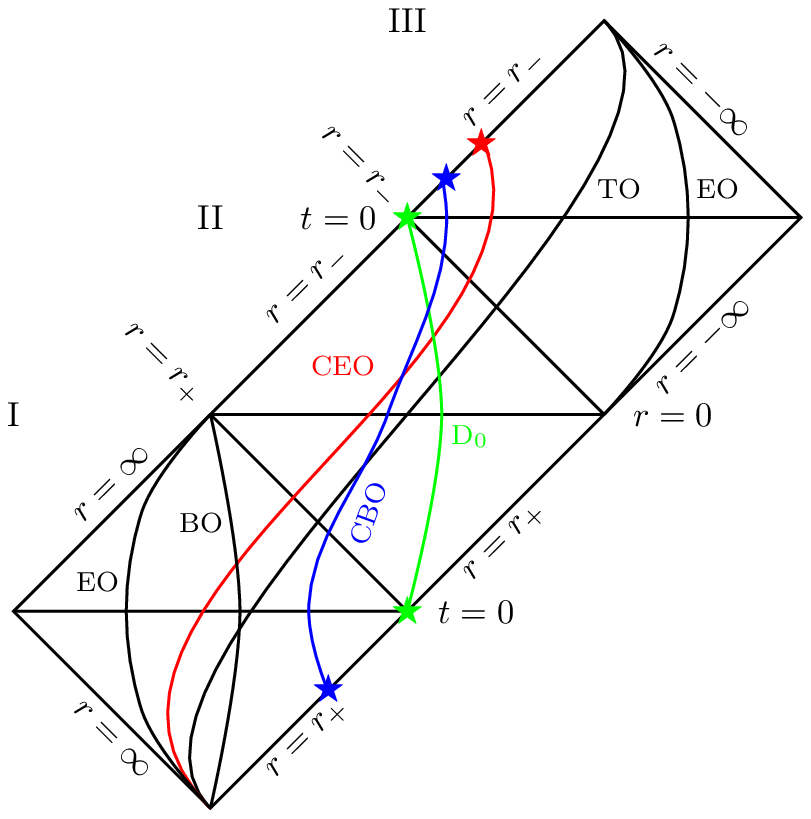}} \qquad
\subfigure[][Taub--NUT space-time II. The CEO (red) starts 
at the horizon $r_-$, the CBO (blue) starts at the horizon $r_-$ 
and terminates at the horizon $r_+$.]{\includegraphics[width=0.45\textwidth]{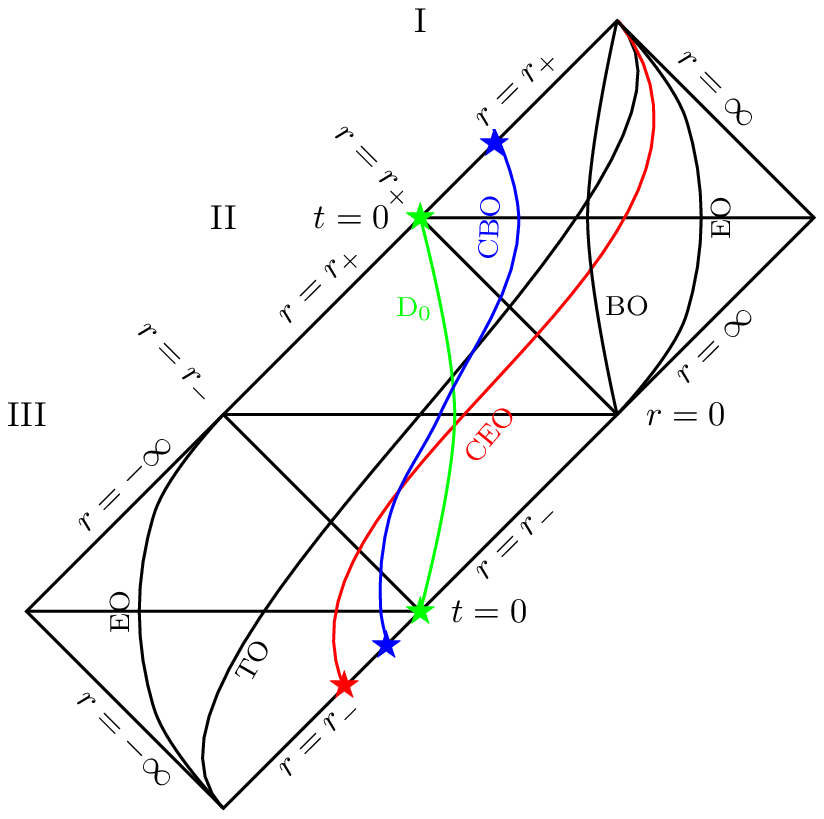}} 
\end{center}
\caption{Topology of orbits in Carter--Penrose diagrams of Taub--NUT space-time. The orbits drawn in black are standard orbits with infinite proper time, the orbits in red, blue, and green are geodesically incomplete. \label{Fig:NUTOrbits} }
\end{figure}

The possible orbits obtained above are depicted in Fig.~\ref{Fig:NUTOrbits}. 
From this diagram it is immediately clear that the CEOs shown in red, 
as well as the CBOs shown in blue (and the ${\rm D}_0$ shown in green 
which represent a special case of the CBOs) 
are geodesically incomplete in a single copy of Taub--NUT space. 
The CEOs, for instance, start at spatial infinity, 
cross the horizons $r_+$ and $r_-$, 
and are forced to end when they reach the horizon $r_-$ a second time.
Therefore, as long as there is only a single copy of Taub--NUT space,
there is no way for such a geodesic to cross the horizon $r_-$ a second time,
implying geodesic incompleteness of the space--time.
This illustrates the problem in the interpretation of Misner and Taub,
where the periodic identification of the time coordinate
precludes the analytic extension of the space--time with further copies.

\subsection{Singularities in the Misner and Taub version 
of the Taub--NUT space--time}\label{ssec:incomplMisner}

The geodesic incompleteness in this version of Taub--NUT space--time 
has been investigated by Misner and Taub \cite{MisnerTaub69}. 
We now explicitly demonstrate this geodesic incompleteness by means of our analytic solution: 
By introducing Eddington--Finkelstein like coordinates 
$\psi_\pm$ through $2 n \psi_{\pm}={t}\pm\int_{r} \frac{\rho^2}{\Dr} d r$,
one eliminates the singular behavior of the metric~\eqref{metrikNUTdeSitter} 
at the horizons ${\Dr}=0$ and one obtains two metrics
\begin{equation}
ds^2=4n^2\frac{\Dr}{\rho^2}\left(d\psi-(\cos\vartheta + C) d\varphi\right)^2 \mp 4n\left(d\psi-(\cos\vartheta + C) d\varphi\right)dr - \rho^2\left(d\vartheta^2+\sin^2\vartheta d\varphi^2\right) \label{newmetric} \ .
\end{equation}
With $I_{\tr}(\gamma)$ and $I_{\vartheta}(\gamma)$ as in~\eqref{tint_NUT} 
the equation for the coordinate $\psi$ takes the form
\begin{eqnarray}
2 \tilde n (\psi_{\pm}-\psi_{\rm in}) &=& (\tilde{t}-\tilde{t}_{\rm in}) \pm \int^{\tr}_{\tr_{\rm in}}{\frac{\tilde\rho^2}{\tDr} d\tr} \nonumber \\  &=&  I_{\tr}(\gamma) + I_{\vartheta}(\gamma) \pm \left( \tr +\half\log |\tDr| +\half \sqrt{1+4 \tilde n^2}\log \Biggl|{\frac{ \frac{2\tr-1}{\sqrt{4 \tilde n^2+1}}-1 }{ \frac{(2\tr-1)}{\sqrt{4 \tilde n^2+1}} +1 } }\Biggr| \right)\Biggl|^{\tr(\gamma)}_{\tr_{\rm in}}  \, . \label{psi1}
\end{eqnarray}

\begin{figure}[t]
\begin{center}
\subfigure[][geodesic incompleteness of Taub--NUT space--time for $\psi_+(\gamma)$]{\label{plusDelta}\includegraphics[width=0.4\textwidth]{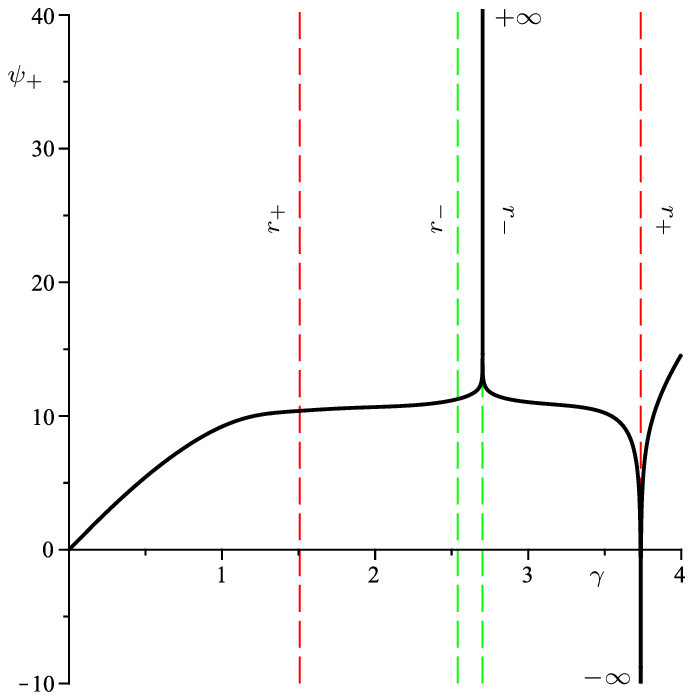}}  \qquad
\subfigure[][geodesic incompleteness of Taub--NUT space--time for $\psi_-(\gamma)$]{\label{minusDelta}\includegraphics[width=0.4\textwidth]{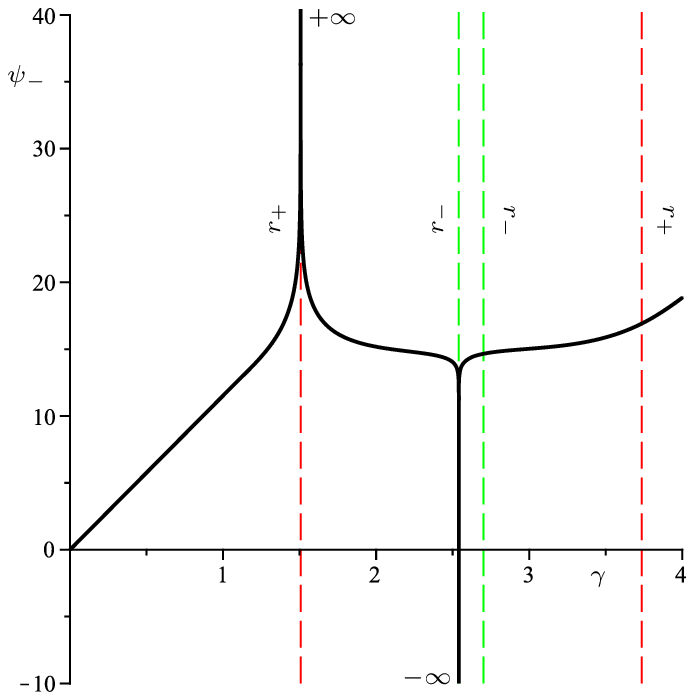}} 
\end{center}
\caption{Taub--NUT space--time: $\psi_{+}$ and $\psi_{-}$ for $\tilde{n}=0.5$, ${k}=1$, $\tilde{L}=2$ and ${E}^2=0.93$. The orbit is a CBO of the type depicted in Fig.~\ref{mu093-term}}   \label{incompleteNUT}
\end{figure}

Figure~\ref{incompleteNUT} shows the $\psi(\gamma)$--dependence 
which reveals the incompleteness of crossover bound geodesics. 
Consider e.g.~$\psi_+$ in Fig.~\ref{plusDelta}. 
A test particle starting its motion somewhere in the region $\tDr>0$ 
can cross the horizons $r_+$ and $r_-$ only once; 
when it approaches $r_-$ for the second time 
the function $\psi(\gamma)$ diverges and prevents 
the particle from leaving the region $\tDr\leq0$. 
This behaviour represents the {\emph{geodesic incompleteness}} 
observed by Misner and Taub~\cite{MisnerTaub69}. 
Also if a particle starts from $\tr=0$ the geodesic is incomplete, 
since once a particle leaves the region $\tDr\leq0$ 
after crossing one of the horizons $r_-$ or $r_+$,
it cannot cross this horizon a second time, 
because $\psi(\gamma)$ diverges there. 
At the same time the affine parameter $\gamma$ cannot be continued 
further, which indicates the geodesic incompleteness
of the Taub--NUT space--time.

Figure~\ref{minusDelta} for $\psi_-$ demonstrates the incompleteness 
of the geodesics at the first pair of horizons.

Thus, the Eddington--Finkelstein--like transformations eliminate 
the singular behaviour of the original Schwarzschild--like coordinates 
only at the first crossing of the horizons,
but lead to incomplete geodesics at the attempted second
crossing.

Because an analytic extension of the space--time 
with a second copy is not possible
without destroying the Hausdorff property of the underlying topological space~\cite{MiKruGo71}, the Taub-NUT space--time with
periodically identified time coordinate is geodesically incomplete.

\subsection{Singularities in the Kruskal--like analytic extension of the Taub--NUT space--time}\label{ssec:incomplKr}

\begin{figure}[th!]
\includegraphics[width=0.5\textwidth]{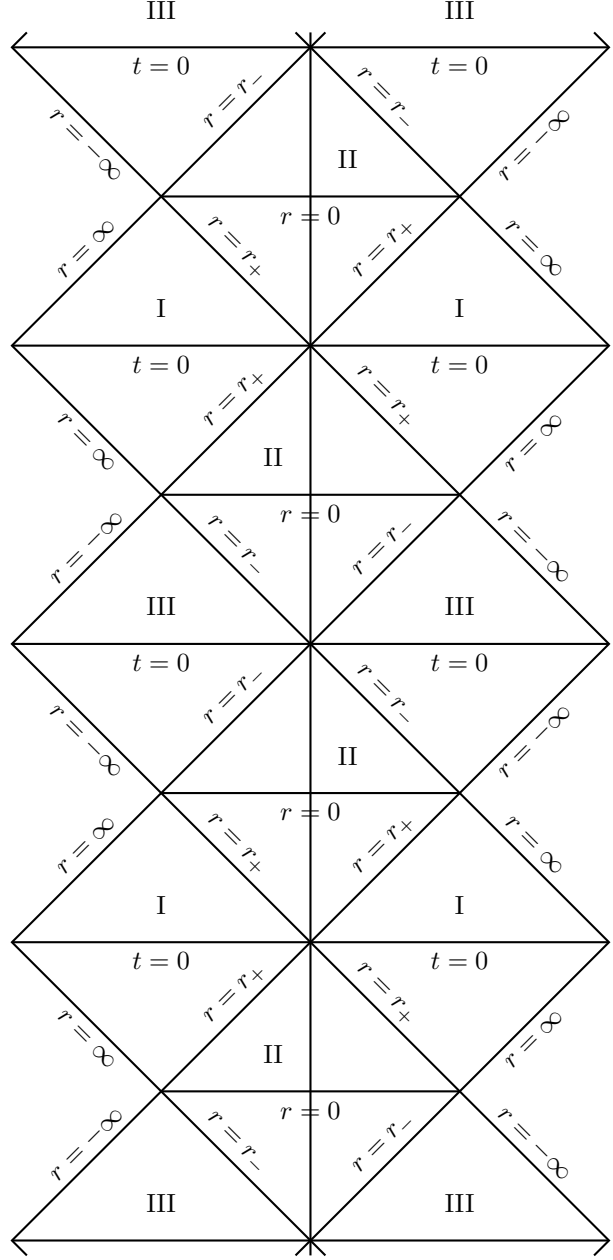}
\caption{Kruskal--like analytic continuation of Taub--NUT space--time 
in a Carter--Penrose diagram. Here region I is given by $r > r_+$, 
region II by $r_- < r < r_+$ and region III by $r < r_-$. 
One copy of Taub--NUT space corresponds to the regions I, II, and III glued together diagonally, 
compare Fig.~\ref{Fig:NUTCP}. \label{Fig:NUTKruskal}}
\end{figure}

In order to extend these incomplete geodesics,
Miller, Kruskal and Godfrey~\cite{MiKruGo71} presented a Kruskal--like 
analytic extension of the Taub--NUT space--time, where an infinite
sequence of further copies of Taub--NUT space--time is added. 
This procedure is completely analogous to the case of the 
Reissner--Nordstr\"om space--time discussed in Appendix~\ref{RN_incompl}.
The Carter--Penrose diagram of this extension of Taub--NUT space--time 
is presented in Fig.~\ref{Fig:NUTKruskal}.

\begin{figure}[th!]
\includegraphics[width=0.5\textwidth]{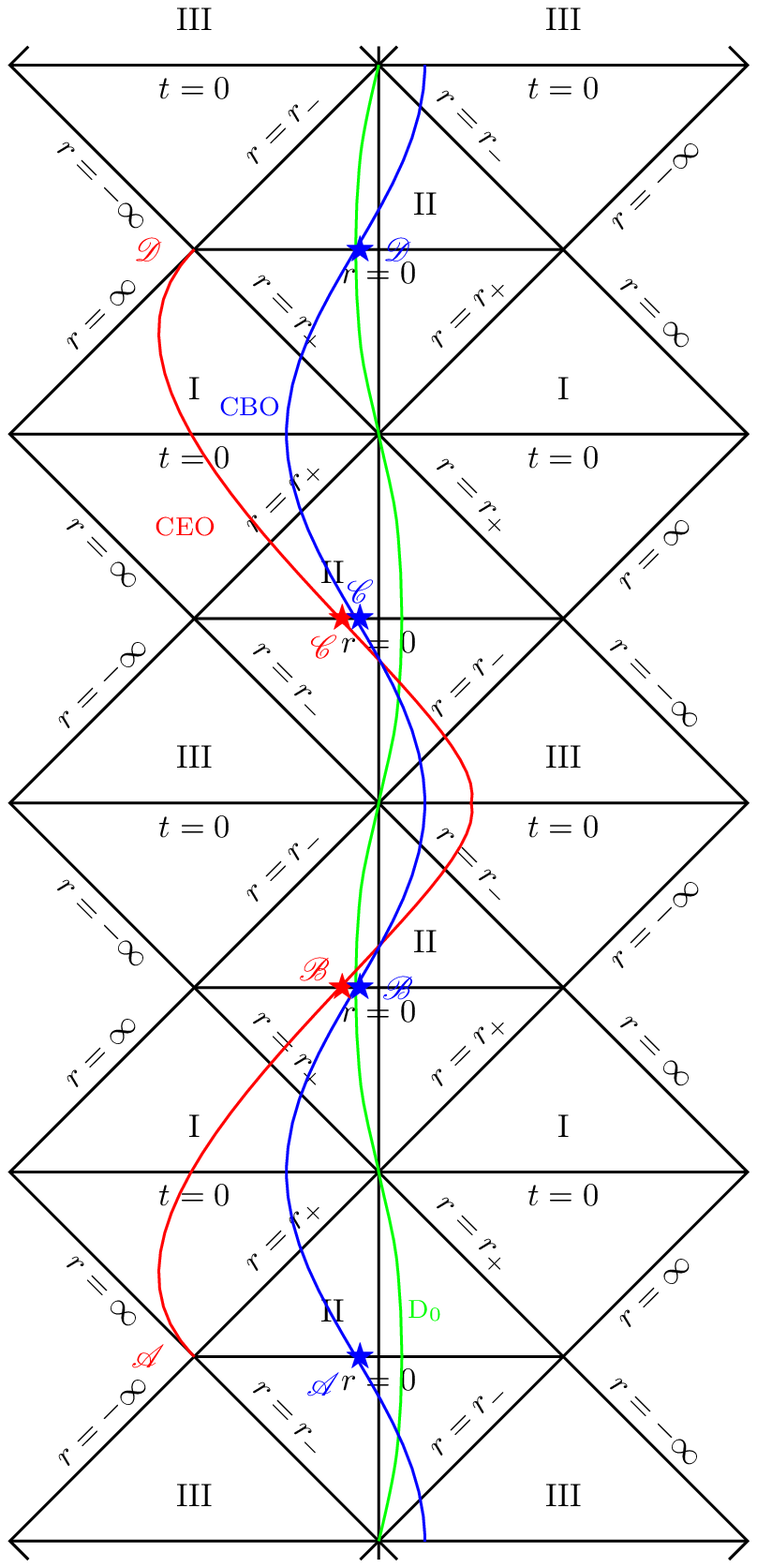}
\caption{The CEO (red) and the CBO (blue)
in the Kruskal--like analytic extension of Taub--NUT space--time.
In the Misner--Taub interpretation of Taub--NUT space--time
these orbits are incomplete at the horizons.
In the Bonnor--Manko--Ruiz interpretation of Taub--NUT space--time
the orbits are incomplete at the singular axis.
The orbits then consist of incomplete pieces:
3 pieces for the CEO (red):
$\mathscr{A}$ $\rightarrow$ $\mathscr{B}$,
$\mathscr{B}$ $\rightarrow$ $\mathscr{C}$,
$\mathscr{C}$ $\rightarrow$ $\mathscr{D}$,
which are incomplete at $\mathscr{B}$ and $\mathscr{C}$; 
an infinite number of pieces for the CBO (blue):
$\mathscr{A}$ $\rightarrow$ $\mathscr{B}$,
$\mathscr{B}$ $\rightarrow$ $\mathscr{C}$,
$\mathscr{C}$ $\rightarrow$ $\mathscr{D}$, etc.,
which are incomplete at $\mathscr{A}$, $\mathscr{B}$, $\mathscr{C}$, $\mathscr{D}$, etc.
Also shown is the orbit ${\rm D}_0$ (green).
\label{Fig:NUTKruskal2}}\end{figure}

In this extended space--time it is possible to continue the CEOs and CBOs. 
These geodesics are therefore no longer forced to terminate at a horizon, 
when trying to pass it for a second time. 
This is illustrated in Fig.~\ref{Fig:NUTKruskal2},
where these orbits are displayed in the Carter--Penrose diagram.

The price to pay for the continuation of the geodesics at the horizons
is, however, that the singular $\vartheta=0, \pi$--axis cannot be eliminated
by a periodic identification of the time coordinate.
Thus, the axis represents a singularity,
where geodesics must terminate.
In the classification of Ellis and Schmidt \cite{EllisSchmidt77,Schmidt}
this type of singularity has been termed a
quasi--regular singularity, since the curvature remains finite
(see also \cite{Geroch,Israel}).
Nevertheless, a geodesic ends when reaching the singularity, 
analogous to the more often encountered conical singularity. 
A further discussion of this type of singularity may be found 
in \cite{Tod} and \cite{Bonnor01}, classifying this 
singularity further as a torsion singularity.

Thus, the introduction of the Kruskal--like analytical extension 
of the space--time eliminates the geodesic incompleteness at the horizons 
but only trades it for a physical singularity along the $\vartheta=0, \pi$--axis,
where the geodesics are forced to terminate: 
consequently the geodesic incompleteness of Taub--NUT space--times 
is retained in either interpretation of this intriguing vacuum solution.

\section{The observables}\label{observables}

As in any other space--time orbits in Taub--NUT space--time
possess some invariantly defined observables like perihelion shift, light deflection, the deflecton angle for flyby orbits, or the Lense--Thirring effect. These quantities are directly or indirectly related to the periods of the $\vartheta$ and the $\tilde r$ motion. 

The two fundamental periods of the $\wp$--function where $2\omega_{1}\in\mathbb{R}$ and $2\omega_{2}\in\mathbb{C}$ are 
\begin{equation}
\omega_{1}=\int^{e_2}_{e_1} \frac{dy}{\sqrt{P_3(y)}} \ , \,\,\,\,\,\, \omega_{2}=\int^{e_3}_{e_2} \frac{dy}{\sqrt{P_3(y)}} \, \ .
\end{equation}
For a bound orbit in the region $(4)_-$ the $\tilde r$--motion of the test particle is periodic $\tr(\gamma+\omega_{\tr})=\tr(\gamma)$, in particular it oscillates between $r_{\rm min}$ and $r_{\rm max}$ with a period $\omega_{\tilde r} = 2 \omega_1$ 
\begin{equation}
\omega_{\tr}= 2 \int_{r_{\rm min}}^{r_{\rm max}} \frac{d\tr}{\sqrt{R}} = 2 \int_{e_1}^{e_{2}} \frac{dy}{\sqrt{P_3(y)}} = 2 \omega_{1} \ ,
\end{equation}
where $e_1$ and $e_{2}$ are the zeros of $P_3(y)$ related to $r_{\rm min}$ and $r_{\rm max}$. The corresponding orbital frequency is $\frac{2\pi}{\omega_{\tr}}$. 

The $\vartheta$--period of a bound orbit in the region $(4)_-$ is simply given by  
\begin{equation}
\omega_{\vartheta}= 2 \int^{\vartheta_{\rm max}}_{\vartheta_{\rm min}} \frac{d\vartheta}{\sqrt{\Theta}}= - 2 \int^{\xi_{\rm min}}_{\xi_{\rm max}} \frac{d\xi}{\sqrt{\Theta_{\xi}}} = \frac{2\pi}{\sqrt{-a}} \ ,
\end{equation}
and the corresponding frequency by $\frac{2\pi}{\omega_{\vartheta}}$.

The secular rates at which the angle $\varphi$ and the time $t$ accumulate are given by (for $\Lp>2{E} \tilde{n}$): 
\begin{equation}
Y_{\varphi} = \frac{2}{\omega_{\vartheta}} \int^{\xi_{{\rm min}}}_{\xi_{{\rm max}}} \frac{\Lp - 2 \tilde n E \xi}{1 - \xi^2} \left(-\frac{d\xi}{\sqrt{\Theta_\xi}} \right)= \frac{1}{\omega_\vartheta} (I_- - I_+)\bigr|^{\xi_{\rm min}}_{\xi_{\rm max}} = \frac{2\pi}{\omega_\vartheta} = \sqrt{-a} \, . 
\end{equation}
and
\begin{eqnarray}
\Gamma & = & \frac{2}{\omega_{\tr}} \int_{r_{\rm min}}^{r_{\rm max}}{E  \frac{\trh^4}{\tDr} \frac{d\tr}{\sqrt{R}}} 
+ \frac{2}{\omega_{\vartheta}} \int_{\vartheta_{\rm min}}^{\vartheta_{\rm max}}{ \left(\Lp - 2 \tilde{n} {E} \cos\vartheta \right) \frac{2 \tilde{n} (\cos\vartheta + C)}{\sin^2\vartheta} \frac{d\vartheta}{\sqrt{\Theta}}  } \nonumber \\ 
 &=& \frac{2}{\omega_{\tr}} I_{\tr}\Bigl|^{\gamma_{e_2} }_{\gamma_{e_1} } + 4\tilde{n}^2E + 2\tilde{n} \sqrt{-a} C\ ,
\end{eqnarray}
where $I_{\tr}$ defined in Eq.~\eqref{IryNUT4} is evaluated at $\gamma_{e_i}$ corresponding to the root $e_i$, $i=1,2$. The orbital frequences $\Omega_r$, $\Omega_\vartheta$ and $\Omega_\varphi$ are then given by:
\begin{equation}
\Omega_{\tr}=\frac{2\pi}{\omega_{\tr}}\frac{1}{\Gamma} \, , \qquad \Omega_\vartheta=\frac{2\pi}{\omega_{\vartheta}}\frac{1}{\Gamma} \, , \qquad \Omega_\varphi=\frac{Y_{\varphi}}{\Gamma} \ .
\end{equation}

As discussed in \cite{DrascoHughes04,FujitaHikida09} the differences between these orbital frequences are related to the perihelion shift and the Lense--Thirring effect  
\begin{eqnarray}
\Delta_{\rm perihelion} & = & \Omega_\varphi - \Omega_{\tr} = \left( \sqrt{-a} - \frac{2\pi}{\omega_{\tr}}  \right)\frac{1}{\Gamma} \\ 
\Delta_{\rm Lense-Thirring} & = & \Omega_\varphi - \Omega_\vartheta = 0 \ .
\end{eqnarray}

Interestingly, there is no Lense-Thirring effect in the Taub--NUT
space-times. The geometrical reason for this is the property
that the orbits lie on cones.
These cones are fixed in space, and so the orbit reaches
the maximal value of $\vartheta$ after
each full revolution of $\varphi$ by $2 \pi$.
Consequently, the frequencies $\Omega_\varphi$
and $\Omega_\vartheta$ coincide.
Thus in Taub--NUT space-times the fixed cone plays the role of the fixed plane of Schwarzschild space-times. We expect that in the case of a Kerr-Taub-NUT space-time the orbital cone in the far field region precesses like the orbital plane precesses in Kerr space-time.

Further possible observables in the Taub--NUT space--time 
are given by the deflection angle for escape and transit orbits 
for massive test particles and for light.

\section{Conclusions and Outlook}

In this paper we presented the analytic solution of the geodesic equation in Taub--NUT space--times in terms of the Weierstra{\ss} $\wp$, $\sigma$ and $\zeta$ functions. The derived orbits depend on the particle's energy, angular momentum, Carter constant and on the parameters of the gravitating source. We discussed the general structure of the orbits and gave a complete classification of their types.

We also addressed the properties of the geodesics
in the light of the interpretations 
of the Taub--NUT metric following on the one hand Misner \cite{Misner63} and
on the other hand Bonnor \cite{Bonnor69}.
According to Misner and Taub~\cite{MisnerTaub69} 
one encounters incomplete geodesics, 
when the affine parameter suddenly terminates 
at the attempt to cross a horizon a second time. 
In their approach Misner and Taub used Eddington--Finkelstein transformations. 
Such kind of transformations lead to incomplete geodesics 
also in the Reissner--Norstr\"om space--time. 
This is due to the shortcoming of these coordinates to not 
provide the complete analytical extension of the space--time. 
In such a Kruskal--like extension of the Taub--NUT space--time 
as performed by Miller, Kruskal and Godfrey~\cite{MiKruGo71} 
the incomplete geodesic behaviour at the horizons is eliminated.
However, other geodesics are then incomplete
in the analytically extended Taub--NUT space--time,
because the singularity at the $\vartheta=0,\pi$ axis is retained.
A periodic identification of the time coordinate
would violate the Hausdorff property of the manifold
and is thus prohibited.
Following Bonnor~\cite{Bonnor69} and
Manko and Ruiz~\cite{Manko2005}
the singularity on the axis must then be considered as a physical singularity.

\section*{Acknowledgement}

We would like to thank B.~Bonnor, D.~Giulini, V.~S.~Manko, E.~Radu and B.~G.~Schmidt 
for helpful discussions. 
E.H. and V.K. acknowledge financial support 
of the German Research Foundation DFG.

\begin{appendix}

\section{Incomplete behavior of geodesics in the Reissner--Nordstr\"om space--times}\label{RN_incompl} 

The $\tr$-- and $\tilde{t}$--equations in the Reissner--Nordstr\"om space--time 
in four dimensions take the form 
(cf.~Eqs.~\eqref{eq-r-theta:1},~\eqref{dtildetdgamma})
\begin{eqnarray}
\left(\frac{d\tr}{d\gamma}\right)^2 & = & R_{\rm RN} \ , \qquad R_{\rm RN}=E^2\tr^4-\Delta_{\rm RN}(\tr^2+\tilde{L}^2) \ , \\
\frac{d\tilde t}{d\gamma} & = & \frac{E\tr^4}{\Delta_{\rm RN}} \ , \qquad \Delta_{\rm RN}=\tr^2-\tr+\eta^2 \ ,
\end{eqnarray}
where $\eta$ is the normalized charge, $\eta=\frac{q}{r_{\rm S}}$,
and the Mino~\cite{Mino03} time $\gamma$ is used
where $\tr^2 d\gamma = d\tilde{\tau}$.
Solutions of these equations can be derived by the same scheme as 
employed in Eq.~\eqref{eq-r-theta:1} with solution~\eqref{solrNUTlight} 
and in Eq.~\eqref{dtildetdgamma} with solution~\eqref{tint_NUT}.
(Note that in the Reissner--Nordstr\"om space--time the motion can be restricted 
to the equatorial plane and only the $I_{\tr}(\gamma)$ part survives.)

\begin{figure}[t]
\begin{center}
\subfigure[][Geodesic incompleteness of a single copy of RN space--time for $\psi_+$]{\label{RN_plusDelta}\includegraphics[width=0.4\textwidth]{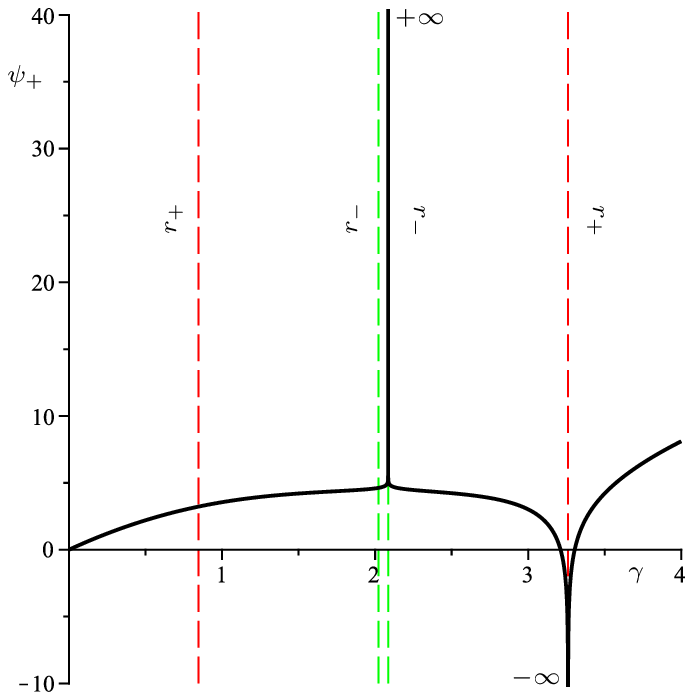}} \qquad
\subfigure[][Geodesic incompleteness of a single copy of RN space--time for $\psi_-$]{\label{RN_minusDelta}\includegraphics[width=0.4\textwidth]{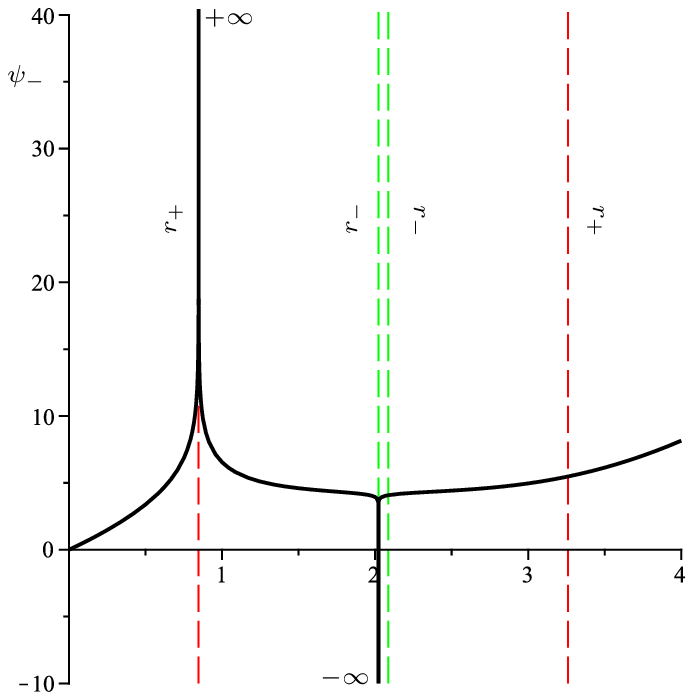}} 
\end{center}
\caption{Reissner--Nordstr\"om space--time: $\psi_{+}$ and $\psi_{-}$ for $\eta=0.4$, $\tilde{L}=2$ and ${E}^2=0.95$.}   \label{incompleteRN}
\end{figure}

The incomplete behaviour of geodesics in the Reissner--Nordstr\"om space--time 
is similar to that in the Taub--NUT space--time. 
Although the orbits are completely regular from the $\tr$--, $\vartheta$-- 
and $\varphi$--equations, this interesting feature reveals itself 
when one investigates the $t$--equation. The Eddington--Finkelstein transformations 
$\displaystyle{\psi_{\pm}
=\tilde{t}\pm\int_{\tr} \frac{\tr^2}{\Delta_{\rm RN}} d\tr}$ convert the metric into a nonsingular form. The affine parameter $\gamma$ stops to increase at some point while $\psi$ diverges there. 
This happens at the attempt to cross one of the horizons a second time, 
as illustrated in Fig.~\ref{incompleteRN}. 
The similarity with Fig.~\ref{incompleteNUT} for the Taub--NUT space--time 
is immediately recognized.

The Kruskal--like extension of the Reissner--Nordstr\"om space--time 
is better suited for the description of the geodesics.
These coordinates present an analytical extension 
of the space-time to the entire range of possible radial and time coordinates, 
and yield an infinite set of copies of the original 
Reissner--Nordstr\"om space--time.
Consequently, the geodesics which seemed to be incomplete at the horizons 
in Eddington--Finkelstein coordinates can be smoothly continued 
through the horizons to the glued regions $\rm I$, $\rm II$, $\rm III$
of the copies. 
(The Carter-Penrose diagram of the Kruskal extension 
for the Reissner--Nordstr\"om space--time can be found,
for instance, in~\cite{Straumann, Chandrasekhar83}.)

\section{Integration of elliptic integrals of the third kind}\label{elldiffIII}

We consider an integral of the type 
$\displaystyle{I_1=\int^v_{v_{\rm in}} \frac{dv}{\wp(v)-p_1}}$. 
This integral is of the third kind because the function $f_1(v)=\left(\wp(v)-p_1\right)^{-1}$ has two simple poles $v_1$ and $v_2$ in a fundamental parallelogram with vertices $0$, $2\omega_1$, $2\omega_1+2\omega_2$, $2\omega_2$, where $2\omega_1$ and $2\omega_2$ are fundamental periods of $\wp(v)$ and $\wp^\prime(v)$. 

Consider the Laurent series for the function $f_1$ around $v_i$
\begin{equation}
f_1(v)= a_{-1,i} (v-v_i)^{-1} + \text{holomorphic part}  \ , \label{f1Laurent}
\end{equation} 
and the Taylor series of $f_1^{-1}$ about $v_i$
\begin{equation}
f^{-1}_1(v)= \wp^\prime(v_i)(v-v_i) + {\cal{O}}(v^2)  \ . \label{f1Taylor}
\end{equation} 
Comparing the coefficients in the equality $1=f_1(v)f^{-1}_1(v)$ where
\begin{equation}
1 = f_1(v)\left(  \wp^\prime(v_i)(v-v_i) + {\cal{O}}(v^2)  \right) =  a_{-1,i} \wp^\prime(v_i) + {\cal{O}}(v^2)   \, \label{f1coeff}
\end{equation} 
yields $\displaystyle{a_{-1,i}=\frac{1}{\wp^\prime(v_i)}}$. Thus, the function $f_1(v)$ has a residue $\frac{1}{\wp^\prime(v_i)}$ in $v_i$.

The Weierstra{\ss} $\zeta(v)$ function is an elliptic function with a simple pole in $0$ and residue $1$. Then the function $A_1=f_1(v)-\sum^{2}_{i=1}\frac{\zeta(v-v_i)}{\wp^\prime(v_i)}$ is an elliptic function without poles and therefore a constant~\cite{Markush}, which can be determined from $f_1(0)=0$. Thus,
\begin{equation}
f_1(v)= \sum^{2}_{i=1}\frac{\zeta(v-v_i)+\zeta(v_i)}{\wp^\prime(v_i)}  \ , \label{f1fin}
\end{equation} 
here $\wp^\prime(v_2)=\wp^\prime(2\omega_j-v_1)=-\wp^\prime(v_1)$.  
Applying now the definition of the Weierstra{\ss} $\sigma$--function $\int^{v}_{v_{\rm in}} \zeta(v)dv=\log\sigma(v)-\log\sigma(v_{\rm in})$ upon the integral $I_1$ we get the solution
\begin{equation}
I_1= \int^v_{v_{\rm in}} f_1(v)dv=\sum^{2}_{i=1}\frac{1}{\wp^\prime(v_i)} \Biggl( \zeta(v_i)(v-v_{\rm in}) + \log\frac{\sigma(v-v_i)}{\sigma(v_{\rm in}-v_i)}  \Biggr)  \ . \label{I1int}
\end{equation}

\section{Integration of elliptic integrals of the type $I_2=\int^v_{v_{\rm in}} \frac{dv}{\left(\wp(v)-p_3 \right)^2}$ }\label{elldiff2}

We consider the Laurent series of $f_2(v)$ and the Taylor series of $f^{-1}_2(v)$ around $v_i$ for $\displaystyle{f_2(v)=\frac{1}{\left(\wp(v)-p_3 \right)^2}}$:
\begin{eqnarray}
  f_2(v) &=& a_{-2,i} (v-v_i)^{-2} + a_{-1,i} (v-v_i)^{-1} + \text{holomorphic part}  \ , \label{f2Laurent} \\
  f^{-1}_2(v) &=& \left( \wp^\prime(v_i)(v-v_i) + \half \wp^{\prime\prime}(v_i) (v-v_i)^2 + {\cal{O}}(v^3) \right)^2 \nonumber \\  & =& \left(\wp^\prime(v_i)\right)^2(v-v_i)^2 + \wp^\prime(v_i) \wp^{\prime\prime}(v_i) (v-v_i)^3 + {\cal{O}}(v^4) \label{f2Taylor} \ . 
\end{eqnarray}
The function $f_2(v)$ has poles of second order in $v_1$ and $v_2$ such that $f_2(v_1)=p_3=f_2(v_2)$. 
Comparison of the coefficients in 
\begin{eqnarray}
1 & = & f_2(v)\left( \left(\wp^\prime(v_i)\right)^2(v-v_i)^2 + \wp^\prime(v_i) \wp^{\prime\prime}(v_i) (v-v_i)^3 + {\cal{O}}(v^4)   \right) \nonumber \\
  & = & a_{-2,i} \left(\wp^\prime(v_i)\right)^2 + (v-v_i) \left[ a_{-1,i}\left(\wp^\prime(v_i)\right)^2 +  a_{-2,i}\wp^\prime(v_i)\wp^{\prime\prime}(v_i)    \right] +  {\cal{O}}(v^2) \label{f2coeff}  \, 
\end{eqnarray} 
yields
\begin{equation}
a_{-2,i}=\frac{1}{\left(\wp^\prime(v_i)\right)^2} \,\, \ , \,\,\,\,\,\,\, a_{-1,i}=-\frac{\wp^{\prime\prime}(v_i)}{\left(\wp^\prime(v_i)\right)^3}  \ . \label{a2a1f2}
\end{equation}

The function $\wp(v)$ possesses a pole of second order in $v=0$ with residuum $0$ and the Laurent series of $\wp$ begins with $v^{-2}$. Then the Laurent series of $a_{-2,i}\wp(v-v_i)$ around $v_i$ begins with $a_{-2,i}(v-v_i)^{-2}$ which is similar to the first term in~\eqref{f2Laurent}. The Laurent series of $a_{-1,i}\zeta(v-v_i)$ around $v_i$ begins with $a_{-1,i}(v-v_i)^{-1}$. Thus, the function $\displaystyle{A_2=f_2(v)-\sum^{2}_{i=1}\left(  \frac{\wp(v-v_i)}{\left(\wp^\prime(v_i)\right)^2} - \frac{\wp^{\prime\prime}(v_i)\zeta(v-v_i)}{\left(\wp^\prime(v_i)\right)^3}  \right)}$ has no poles and is constant~\cite{Markush} and can be calculated from $f_2(0)=0$: $\displaystyle{A_2=-\sum^{2}_{i=1}\left(  \frac{\wp(v_i)}{\left(\wp^\prime(v_i)\right)^2} + \frac{\wp^{\prime\prime}(v_i)\zeta(v_i)}{\left(\wp^\prime(v_i)\right)^3}  \right)}$.

Using of $\int^v_{v_{\rm in}}\wp(v)dv = - \zeta(v) + \zeta(v_{\rm in})$ and the definition of the $\sigma$--function the integral $I_2$ takes the form:
\begin{equation}
I_2 = \int^v_{v_{\rm in}} f_2(v)dv= A_2 (v-v_{\rm in})  
- \sum^{2}_{i=1} \Biggl[ \zeta(v-v_i) - \zeta(v_{\rm in}-v_i) + \frac{\wp^{\prime\prime}(v_i)}{\wp^\prime(v_i)}
 \log\frac{\sigma(v-v_i)}{\sigma(v_{\rm in}-v_i)} \Biggr]\frac{1}{\left(\wp^\prime(v_i)\right)^2}   \ . \label{I2int}
\end{equation}

\end{appendix}


\bibliographystyle{unsrt}

\end{document}